\documentclass[12pt]{article}
\usepackage{amssymb}
\usepackage{graphicx}

\font\gg=eufm10 at 12pt
\def\goth#1{\hbox{\gg #1}}
\def\bthm#1{\vskip10pt\noindent{\bf#1}\space\bgroup\em}
\def\ethm{\egroup\vskip10pt}
\def\rem{\vskip10pt\noindent}
\def\bbox{\mbox{\large\lower.3ex\hbox{\large$\Box$}}}
\def\forall{\hbox{ for all }}
\let\dsty\displaystyle

\begin{document}
\pagestyle{empty}
\vspace* {13mm}
\baselineskip=24pt
\begin{center}
{\bf DETERMINANTS AND INVERSION OF GRAM MATRICES IN FOCK REPRESENTATION
OF $\{ q_{kl}\} $- CANONICAL COMMUTATION RELATIONS AND APPLICATIONS TO
HYPERPLANE ARRANGEMENTS AND QUANTUM GROUPS. PROOF OF AN EXTENSION OF ZAGIER'S
CONJECTURE}
\\[20mm]

S. Meljanac$^1$ and D. Svrtan$^2$ \\[15mm]
{\it $^1$ Rudjer Boskovi\'c Institute - Bijeni\v cka c. 54, 10000 Zagreb,
Croatia \\[5mm]
$^2$ Dept. of Math., Univ. of Zagreb, Bijeni\v cka c. 30, 10000 Zagreb, Croatia}
\end{center}

\newpage
\tableofcontents
\newpage
\begin{center}
{\Large\bf Introduction}\addcontentsline{toc}{section}{Introduction}
\end{center}
\vskip 1cm

Following Greenberg, Zagier, Bo\v zejko and Speicher and others we study
a collections of operators $a(k)$ satisfying the "$q_{kl}$-canonical
commutation relations "
$$a(k)a^{\dagger}(l) - q_{kl}a^{\dagger}(l)a(k) = \delta_{kl}$$
(corresponding for $q_{kl}=q$ to Greenberg (infinite) statistics, for
$q=\pm 1$ to classical Bose and Fermi statistics). We show that $n!\times n!$
matrices $A_{n}(\{ q_{kl}\} )$ representing the scalar products of n-particle states is positive definite for all n if $|q_{kl}|<1$, all k,l ,so that the above commutation relations have a Hilbert space realization in this case. This is
achieved by explicit factorizations of $A_{n}(\{ q_{kl}\} )$ as a product of
matrices of the form $(1-QT)^{\pm 1}$, where Q is a diagonal matrix and T is
a regular representation of a cyclic matrix. From such factorizations we
obtain in Theorem 1.9.2 explicit formulas for the determinant of $A_{n}(\{ q_{kl}\} )$ in the
generic case (which generalizes Zagier's 1-parametric formula). The problem of
computing the inverse of $A_{n}(\{ q_{kl}\} )$ in its original form is
computationally intractable (for $n=4$ one has to invert a $24\times 24$
symbolic matrix). Fortunately, by using another approach (originated by
Bo\v zejko and Speicher ) we obtain in Theorem 2.2.6 a definite answer to that inversion problem
in terms of maximal chains in so called subdivision lattices. Our algorithm in Proposition 2.2.18
for computing the entries of $A_{n}(\{ q_{kl}\} )$ is very efficient. In particular
for $n=8$, when all $q_{kl}=q$, we found a counterexample to Zagier's
conjecture
concerning the form of the denominators of the entries in the inverse of
$A_{n}(q)$. In Corollary 2.2.8 we formulate and prove Extended Zagier's
Conjecture which turns to be the best possible in the multiparametric case
and which implies in one parametric case an interesting extension of the
original Zagier's Conjecture.

By applying a faster algorithm in Proposition 2.2.19 we obtain in Theorem
2.2.20 explicit formulas for the inverse of the matrices $A_n(\{q_{kl}\})$ in
the generic case.

Finally, there are applications of the results above to discriminant
arrangements of hyperplanes and to contravariant forms of certain quantum
groups. \vskip20pt

{\em Acknowledgement.} We would like to thank Prof. Richard Stanley and Prof. Phil
Hanlon for bringing to our attention some papers of Varchenko on quantum
bilinear forms.

\newpage
\setcounter{page}{1}
\pagestyle{plain}
\def\leer{\vspace{5mm}}
\setcounter{equation}{0}
\section{Multiparametric quon algebras, Fock-like representations and
determinants} 
\subsection{$q_{ij}$-canonical commutation relations} 

Let ${\bf q} = \{ q_{ij} : i,j \in I,\bar{q}_{ij}=q_{ji} \} $ be a
 hermitian family of complex numbers (parameters), where I is a finite
(or infinite) set of indices.

By a {\em multiparametric quon algebra} ${\cal A} ={\cal A}^{({\bf q})}$ we shall mean
an associative (complex) algebra generated by $\{ a_{i}, a_{i}^{\dagger},
i \in I \}$ subject to the following $q_{ij}$- canonical commutation relations
\begin{equation}
a_{i}a_{j}^{\dagger} = q_{ij}a_{j}^{\dagger}a_{i} + \delta_{ij},\ \
 \ \forall i,j \in I
\end{equation}
Shortly, we shall give an explicit Fock-like representation of the algebra
${\cal A}^{({\bf q})}$ on the free associative algebra ${\bf f}$
(the algebra of
noncommuting polynomials in the indeterminates $\theta _{i}, i \in I$)
with $a_{i}$ acting
as a generalized $q_{ij}$-deformed partial derivative ${}_{i}\partial =
{}_{i}^{{\bf q}}\partial$ w.r.t. the variable $\theta _{i}$ (the i-th annihilation operator),
and
$a_{i}^{\dagger}$ as multiplication by $\theta _{i}$
(the i-th creation operator). Moreover $a_{i}^{\dagger}$ will be adjoint to
$a_{i}$ w.r.t. a certain sesquilinear form $(\ ,\ )_{{\bf q}}$
on ${\bf f}$ which will be better described via a certain canonical ${\bf q}$-deformed bialgebra
structure on ${\bf f}$, generalizing the one used by Lusztig in his excellent
treatment of quantum groups [Lus]. Then by explicit computation (which extends
Zagier's method) of the determinant of $(\ ,\ )_{{\bf q}}$ we show that
$(\ ,\ )_{{\bf q}}$
is positive definite provided the following condition on the parameters
$q_{ij}$ holds true :
\begin{equation}
|q_{ij}| < 1,\ \ \forall i,j \in I
\end{equation}
The condition (2) ensures that all the many-particle states
$a_{i_{1}}^{\dagger}\cdots a_{i_{r}}^{\dagger}|0> = \theta _{i_{1}}
\cdots \theta _{i_{r}}$,\ $i_{j} \in I, r \geq 0$,
are linearly independent, so we obtain a Hilbert space realization
of the $q_{ij}$-canonical commutation relations (1).

We first need some notations:
\begin{description}
\item[\hbox{${\bf N}=$}] $\{ 0,1,2,\dots \} =$ the set of nonnegative integers
\item[\hbox{$({\bf N}[I],+) =$}] the {\em weight monoid} i.e. the set of all
finite formal linear
                            combinations $\nu = \sum _{i \in I}\nu _{i}i$,
                            $\nu _{i}\in {\bf N}, i\in I$ with componentwise addition
                            $\nu + \nu^{'} = \sum_{i\in I}( \nu _{i} + \nu _{i}
                             ^{'})i$
\item[$|\nu |=$] $\sum_{i\in I}\nu _{i}  \in {\bf N}$ for $\nu = \sum_{i\in I}
\nu _{i}i  \in {\bf N}[I]$
\end{description}

Sometimes it is customary to view the elements $\nu =\sum \nu_{i}i$
of ${\bf N}[I]$ as multisets M in which $i$ appears $\nu_{i}$ times
(in case $\nu_{i}\leq 1$ we have sets contained in $I$) and then $+$
corresponds to the union of multisets and $|\nu |$ is just the cardinality
of M.
\begin{description}
\item[$\beta :$] $({\bf N}[I],+)\times ({\bf N}[I],+) \longrightarrow ({\bf C},
\cdot )$,
the bilinear form on $({\bf N}[I],+)$ with values in the multiplicative monoid
of complex numbers given by $i,j \mapsto q_{ij}$, i.e.
for $\nu = \sum_{i\in I}\nu _{i}i$, $\nu^{'} = \sum_{j\in I}
\nu _{j}^{'}j$, $\beta (\nu ,\nu^{'}) = \prod_{ij}q_{ij}
^{\nu _{i}\nu _{j}^{'}}$.
\end{description}

\subsection{The algebra ${\bf f}$ } 

We denote by ${\bf f}$ the free associative ${\bf C}$-algebra with generators
$\theta _{i}(i\in I)$. For any weight $\nu = \sum_{i\in I}\nu _{i}i \in
{\bf N}[I]$ we denote by ${\bf f}_{\nu }$ the corresponding {\em weight space},
i.e. the subspace of ${\bf f}$ spanned by monomials
$\theta _{{\bf i}} = \theta _{i_{1}}\cdots \theta _{i_{n}}$ indexed by
sequences ${\bf i}=i_{1}\dots i_{n}$ of weight $\nu $,$|{\bf i}|=\nu $
(this means that the number of occurrences of $i$ in ${\bf i}$ is equal
to $\nu _{i}, \forall i \in I$). Then each ${\bf f}_{\nu }$ is a finite
dimensional complex vector space and we have a direct sum decomposition
${\bf f} = \bigoplus
_{\nu }{\bf f}_{\nu }$, where $\nu $ runs over ${\bf N}[I]$. We have
${\bf f}
_{\nu }{\bf f}_{\nu ^{'}}
\subset {\bf f}_{\nu +\nu ^{'}}$, $1\in {\bf f}_{0}$ and $\theta _{i}\in
{\bf f}_{(i)}$.
An element $x$ of ${\bf f}$ is said to be {\em homogeneous} if it belongs to
${\bf f}_{\nu }$
for some $\nu $. We than say that $x$ has {\em weight} $\nu $ and write $|x|=\nu $.\\
We shall consider the tensor product ${\bf f}\otimes {\bf f}$ with the
following $q_{ij}$-{\em deformed multiplication}
\begin{eqnarray*}
(x_{1}\otimes x_{2})(x_{1}^{'}\otimes x_{2}^{'})&=&\beta (|x_{2}|,
|x_{1}^{'}|)x_{1}x_{1}^{'}\otimes x_{2}x_{2}^{'}\\
&=&(\prod_{i,j}q_{ij}^{\nu_{i}\nu_{j}^{'}})x_{1}x_{1}^{'}\otimes
x_{2}x_{2}^{'}, \hbox{ if }x_{2}\in {\bf f}_{\nu }, \ x_{1}^{'}\in {\bf
f}_{\nu^{'}}
\end{eqnarray*}
where $x_{1}, x_{1}^{'}, x_{2}, x_{2}^{'}\in {\bf f}$ are homogeneous; this
algebra is associative since $\beta (\nu ,\nu^{'})$ is bilinear.

The following statement is easily verified: if $r=r_{{\bf q}}
: {\bf f}
\longrightarrow
{\bf f}\otimes {\bf f}$ is the unique algebra homomorphism such that
$r(\theta _{i}
)=\theta_{i}\otimes 1+1\otimes \theta_{i}, \forall i$, then $(r\otimes
1)r=(1\otimes r)r$ takes the same value on any algebra generator $\theta_{i}$,
namely $\theta_{i}\otimes 1\otimes 1 + 1\otimes \theta_{i}\otimes 1 +
1\otimes 1\otimes \theta_{i}$ yielding the coassociativity
property. Thus the algebra ${\bf f}$ with the comultiplication $r$ is an
example of a $q_{ij}$-{\em deformed bialgebra}.

Note that
\begin{eqnarray*}
r(\theta_{i}\theta_{j})&=&r(\theta_{i})r(\theta_{j})=(\theta_{i}\otimes 1
+1\otimes \theta_{i})(\theta_{j}\otimes 1+1\otimes \theta_{j})\\
&=&\theta_{i}\theta_{j}\otimes 1+q_{ij}\theta_{j}\otimes \theta_{i}+
\theta_{i}\otimes \theta_{j}+1\otimes \theta_{i}\theta_{j}
\end{eqnarray*}
More generally we have the following explicit formula for the value
of $r$ on a monomial $\theta_{{\bf i}}=\theta_{i_{1}}\theta_{i_{2}}
\cdots \theta_{i_{n}}$:
$$r(\theta_{{\bf i}})=\sum_{k+l=n, g=(k,l)-shuffle}q_{{\bf i},g}
\theta_{i_{g(1)}}\cdots \theta_{i_{g(k)}}\otimes \theta_{i_{g(k+1)}}
\cdots \theta_{i_{g(k+l)}}$$
where a $(k,l)-shuffle$ is a permutation $g\in S_{k+l}$ such that
$g(1)<g(2)<\cdots <g(k)$ and $g(k+1)<g(k+2)<\cdots <g(k+l)$
and where for $g\in S_{n}$ we denote by $q_{{\bf i},g}$the quantity
$$q_{{\bf i},g}:=\prod_{a<b,g(a)>g(b)}q_{i_{a}i_{b}}$$

\subsection{The sesquilinear form $(\ ,\ )_{{\bf q}}$ on ${\bf f}$ } 

Note that $r$ maps ${\bf f}_{\nu }$ into $\bigoplus_{(\nu^{'}+\nu^{''}=\nu )}
{\bf f}_{\nu^{'}}\bigotimes {\bf f}_{\nu^{''}}$. Then the linear maps
${\bf f}_{\nu^{'}+\nu^{''}}\longrightarrow {\bf f}_{\nu^{'}}\bigotimes
{\bf f}_{\nu^{''}}$ defined by $r$ give,
by passage to dual spaces, linear maps ${\bf f}_{\nu^{'}}^{*}\bigotimes
{\bf f}_{\nu^{''}}^{*}\longrightarrow {\bf f}_{\nu^{'}+\nu^{''}}^{*}$.
These define the structure of an associative algebra with 1 on $\bigoplus
_{\nu }{\bf f}_{\nu }^{*}$.
For any $i\in I$, let $\theta_{i}^{*}\in {\bf f}_{i}^{*}$ be the linear
form given by $\theta_{i}^{*}(\theta_{j})=\delta_{ij}$. Let $\Phi :
{\bf f}\longrightarrow \bigoplus_{\nu }{\bf f}_{\nu }^{*}$ be the unique
conjugate-linear algebra homomorphism preserving 1, such that
$\Phi (\theta_{i})=
\theta_{i}^{*},\forall i$.
For $x,y \in {\bf f}$, we set
$$(x,y)_{{\bf q}} = \Phi (y)(x)$$
Then $(\ ,\ )=(\ ,\ )_{{\bf q}}$ is a unique sesquilinear form on ${\bf f}$
such that

{\bf a)}  $(\theta_{i},\theta_{j})=\delta_{ij}, \  \forall i,j \in I$

{\bf b)}  $(x,y^{'}y^{''})=(r(x),y^{'}\bigotimes y^{''}), \  \forall x,y^{'},y^{''}
\in {\bf f}$

{\bf c)} $(xx^{'},y^{''})=(x\bigotimes x^{'},r(y^{''})), \
\forall x,x^{'},y^{''} \in {\bf f}$

\noindent(The sesquilinear form $({\bf f}\bigotimes {\bf f})\times ({\bf f}\bigotimes
{\bf f})\longrightarrow
{\bf C}$ given by $x_{1}\bigotimes x_{2}, x_{1}^{'}\bigotimes x_{2}^{'}
\longrightarrow (x_{1},x_{1}^{'})(x_{2},x_{2}^{'})$ is denoted again by
$(\ ,\ )$). Clearly,

{\bf d)}  $(x,y)=0$ if $x$ and $y$ are homogeneous with $|x|\neq |y|$. In
particular,
        the subspaces ${\bf f}_{\nu }, {\bf f}_{\nu^{'}}$ are orthogonal w.r.t.
$(\ ,\ )$
        for $\nu \neq \nu^{'}$.

{\bf e)}  Let $\rho : {\bf f}\longrightarrow {\bf f}$ be the antiautomorphism of
algebras
with 1 which takes $\theta_{i}$ to $\theta_{i}$
(thus $\rho (\theta_{i_{1}}\cdots \theta_{i_{n}})=\theta_{i_{n}}
\cdots \theta_{i_{1}}$). Then $(\rho (x),\rho (x^{'}))=(x,x^{'}), \
\forall x,x^{'} \in {\bf f}$.

\subsection{The $q_{ij}$-deformed partial derivative maps $_{i}^{{\bf q}}
\partial $ and $^{{\bf q}}\partial_{i}$ } 

Let $i\in I$. Clearly there exists a unique ${\bf C}$-linear map
${}_{i}\partial ={}_{i}^{{\bf q}}\partial :{\bf f}\longrightarrow {\bf f}$ such
that
${}_{i}\partial (1)=0$, ${}_{i}\partial (\theta_{j})=\delta_{ij}$, \ $\forall j$
and obeying the generalized Leibniz rule :

{\bf a)}  $_{i}\partial (xy)={}_{i}\partial (x)y + \beta (i,|x|)x _{i}\partial (y)$                 $ ={}_{i}\partial (x)y + \prod_{j}q_{ij}^{\nu_{j}}x_{i}\partial (y)$, if $x\in {\bf f}_{\nu }$\\
for all homogeneous $x$,$y$. If $x\in {\bf f}_{\nu }$ we have $_{i}
\partial (x)
\in {\bf f}_{\nu -i}$ if $\nu_{i} \geq 1$ and $_{i}\partial (x)=0$
if $\nu_{i}=0$;
 moreover
$r(x)=\theta_{i}\bigotimes {}_{i}\partial (x)$ $ +
$ terms of other bihomogeneities.

Similary, there is a unique ${\bf C}$-linear map
$\partial_{i}={}^{{\bf q}}\partial_{i} : {\bf f}\longrightarrow {\bf f}$
such that
$\partial_{i}(1)=0$,\ $\partial_{i}(\theta_{j})=\delta_{ij}$
for all j and $\partial_{i}(xy)=\beta (|y|,i)\partial_{i}(x)y +
x\partial_{i}(y)$ $(=(\prod_{j}q_{ji}^{\nu_{j}})\partial_{i}(x)y +
x\partial_{i}(y)$, if $y\in {\bf f}_{\nu }$) for all homogeneous $x$,$y$.
If $x\in f_{\nu }$ we have $\partial_{i}(x)\in f_{\nu -i}$ if
$\nu_{i} \geq 1$ and $\partial_{i}
(x)=0$ if $\nu_{i}=0$; moreover, $r(x)=\partial_{i}(x)\bigotimes
\theta_{i} +$ terms
of other bihomogeneities.

>From the definition we see that

{\bf b)}  $(\theta_{i}y,x)=(y,{}_{i}\partial (x)), (y\theta_{i},x)=
(y,\partial_{i}(x))$,
        for all $x,y$ \\
i.e. the operator ${}_{i}\partial $ (resp. $\partial_{i}$) is the adjoint of left
(resp. right) multiplication by $\theta_{i}$.

{\bf c)}  $\rho \partial_{i}={}_{i}\partial \rho $ ($\Rightarrow \partial_{i}=
\rho _{i}\partial \rho^{-1}$)\\
We shall need the following explicit formula for ${}_{i}\partial ={}_{i}^
{{\bf q}}\partial :
{\bf f}\longrightarrow {\bf f}$

{\bf d)}  ${}_{i}\partial (\theta_{j_{1}}\cdots \theta_{j_{n}})=
\sum_{(p:j_{p}=i)}q_{ij_{1}}\cdots q_{ij_{p-1}}\theta_{j_{1}}\cdots
\hat{\theta }_{j_{p}}\cdots \theta_{j_{n}}$\\
where $\hat{}$ denotes omission of the factor $\theta_{j_{p}}$. This formula is obtained by iterating the recursive definition a) for $_{i}\partial $ or by
using the general formula for $r$ in 1.2.
A similar formula holds for $\partial_{i}$.

{\bf e)}  Finally, we note that the form $(\ ,\ )=(\ ,\ )_{{\bf q}}$ will be
nondegenerate
if either of the following conditions holds: let $x\in {\bf f}_{\nu }$, where
$\nu\in {\bf N}[I]$ is different from 0\\
$\alpha $)  If $_{i}\partial (x)=0, \ \forall i$, then $x=0$\\
$\beta $)   If $\partial_{i}(x)=0, \ \forall i$, then $x=0$.

\subsection{Fock-like representations of the multiparametric quon algebra
${\cal A}^{({\bf q})}$ } 

Here we give a representation of the multiparametric quon
algebra ${\cal A}={\cal A}^{({\bf q})}$ (defined in 1.1) on the underlying
vector space of the free associative algebra ${\bf f}$.
\bthm{PROPOSITION 1.5.1.} For each $i\in I$ let $a_{i}^{\dagger}$ act on ${\bf f}$
 as left
multiplication by $\theta_{i}$ and let $a_{i}$ act as the linear map $_{i}
\partial $ defined in 1.4. Then \\
a)  $a_{i}, a_{i}^{\dagger}$ make ${\bf f}$ into a left ${\cal A}$ - module\\
b)  $a_{i}^{\dagger}$ is adjoint to $a_{i}$ w.r.t. the sesquilinear form
$(\ ,\ )=(\ ,\ )_{{\bf q}}$ defined in 1.3.\\
c)  $a_{i} : {\bf f} \longrightarrow {\bf f}$ is locally nilpotent for every
$i\in I$.
\ethm
{\bf Proof.} a) The identity $a_{i}a_{j}^{\dagger}=q_{ij}a_{j}^{\dagger}a
_{i} + \delta_{ij}$ (as maps ${\bf f}\longrightarrow {\bf f}$) follows from
the following computation:\\
${}_{i}\partial (\theta_{j}y)={}_{i}\partial (\theta_{j})y + \beta (i,|
\theta_{j}|)\theta_{j}\ {}_{i}\partial (y)$\\
$=\delta_{ij}y + q_{ij}\theta_{j}\ {}_{i}\partial (y)$ (because $|\theta_{j}|
=j \in {\bf N}[I]$)

For b) see 1.4 b).

c) If $x\in {\bf f}_{\nu }$, then $a_{i}(x)\in {\bf f}_{\nu -i}$ if
$\nu_{i}\geq 1$
and $a_{i}(x)=0$ if $\nu_{i}=0$. It follows that $a_{i} :
{\bf f}\longrightarrow {\bf f}$
is locally nilpotent. The proposition is proved.\\
(Observe that the property $_{i}\partial (1)=0$ is just the vacuum condition
for $a_{i}$, with $1\in {\bf f}$ playing the role of the vacuum vector $|0>$.
)

\subsection{The matrix $A({\bf q})$ of the sesquilinear form
$(\ ,\ )_{{\bf q}}$ on ${\bf f}$ } 

Here we study the sesquilinear form $(\ ,\ )_{{\bf q}}$ on ${\bf f}$,
defined in 1.2, via the associated matrix w.r.t. the basis $B=\{ \theta_{{\bf i}}=\theta_{i_{1}}
\cdots \theta_{i_{n}} | i_{j}\in I, n\geq 0\} $ of the complex vector space
${\bf f}=\bigoplus_{\nu }{\bf f}_{\nu }$. Let $B^{'}=\{ \theta_{{\bf i}}=
\theta_{i_{1}}
\cdots \theta_{i_{n}} | i_{1},\dots ,i_{n}$ all distinct$\} $ and $B^{''}=
B\setminus B^{'}=
\{ \theta_{i_{1}}\cdots \theta_{i_{n}} |$ not all $ i_{1},\dots ,i_{n}$
distinct$\} $. Then we have the direct sum decomposition
\begin{equation}
{\bf f} = {\bf f}^{'}\bigoplus {\bf f}^{''},\ \ \  where \ \
{\bf f}^{'}=span B^{'},\ \ \
{\bf f}^{''}=span B^{''}
\end{equation}
Note that for any weight $\nu =\sum \nu_{i}i \in {\bf N}[I]$ we have
${\bf f}_{\nu }\subset {\bf f}^{'}$ (resp.${\bf f}_{\nu }\subset
{\bf f}^{''}$) if all $\nu_{i}\leq 1$
(resp. some $\nu_{i}\geq 2$). Then we call such weight $\nu $ {\em generic}
(resp. {\em degenerate} ) and we have further direct sum decompositions
\begin{equation}
{\bf f}^{'} = \bigoplus_{\nu \ generic}{\bf f}_{\nu },\ \ \ {\bf f}^{''} =
\bigoplus_{\nu \ degenerate}{\bf f}_{\nu }
\end{equation}

\bthm{PROPOSITION 1.6.1.} i) Let ${\bf A}={\bf A}({\bf q}) : {\bf f}
\longrightarrow {\bf f}$ be the linear operator,
associated to the sesquilinear form $(\ , \ )=(\ ,\ )_{{\bf q}}$ on ${\bf f}$
defined by
$${\bf A}(\theta_{{\bf j}}) = \sum_{{\bf i}}(\theta_{{\bf j}},
\theta_{{\bf i}})_{{\bf q}}\theta_{{\bf i}}$$
Then the ${\bf f}^{'}, {\bf f}^{''}, {\bf f}_{\nu }\ (\nu \in {\bf N}[I])$
are all invariant subspaces
of ${\bf A}$, yielding the following block decompositions
for the corresponding matrices
$$A=A^{'}\bigoplus A^{''}, A^{'}=\bigoplus_{\nu \ \rm generic}A^{(\nu )}, A^{''}=
\bigoplus_{\nu \ \rm degenerate}A^{(\nu )}$$
Moreover, for the matrix entries we have the following formulas:

ii) Let ${\bf i}=i_{1}\dots i_{n}$ and ${\bf j}=j_{1}\dots j_{n}$ be any two sequences with the same generic weight $\nu $ and let $\sigma =
\sigma ({\bf i,j}) \in S_{n}$ be the unique permutation
such that $\sigma \cdot {\bf i}={\bf j}$ (i.e. $i_{\sigma ^{-1}(p)}=
j_{p}$, all $p$). Then
$$A^{'}_{{\bf i,j}} = A^{(\nu )}_{{\bf i,j}} = q_{{\bf i},\sigma }
(=\bar{q}_{{\bf j},\sigma ^{-1}})$$
where (c.f. 1.2)
$$q_{{\bf i},\sigma }:=\prod_{(a,b)\in I(\sigma )}q_{i_{a}i_{b}}$$
with $I(\sigma )=\{ (a,b) | a<b, \sigma (a)>\sigma (b)\} $ denoting the set of inversions of $\sigma $.

iii) Let ${\bf i}=i_{1}\dots i_{n}$ and ${\bf j}=
j_{1}\dots j_{n}$
 be any two sequences of the same degenerate weight $\nu $ and let
${ \bf \sigma  } ({\bf i,j})=\{ \sigma \in S_{n} |
 i_{\sigma^{-1} (p)}=j_{p}$, all $p\} $. Then
$$A^{''}_{{\bf i,j}}=A^{(\nu )}_{{\bf i,j}}=\sum_{\sigma \in
{ \bf \sigma  } ({\bf i,j})}q_{{\bf i},\sigma^{-1}} \ (=\sum_{\sigma \in
{ \bf \sigma  } ({\bf i,j})}\bar{q}_{{\bf j},\sigma^{-1} }).$$
\ethm
{\bf Proof.}\ \ i) follows from 1.3d)

ii) We have, by 1.4b)
$$A^{'}_{{\bf i,j}}=A_{{\bf i,j}}=(\theta_{{\bf j}},\theta_{{\bf i}})_{{\bf q}}=
({}_{i_{1}}\partial (\theta_{{\bf j}}),\theta_{i_{2}}\cdots
\theta_{i_{n}})_{{\bf q}}=\cdots ={}_{i_{n}}\partial \cdots {}_{i_{1}}
\partial (\theta_{j_{1}}\cdots \theta_{j_{n}})$$
Now by applying the formula 1.4d) successively for $i=i_{1},i_{2},\dots $
and if $j_{\sigma (1)}=i_{1}, j_{\sigma (2)}=i_{2},\cdots $ we obtain
$$(\prod_{1<b, \sigma (b)<\sigma (1)}q_{i_{1}i_{b}})(\prod_{2<b, \sigma (b)<
\sigma (2)}q_{i_{2}i_{b}})\cdots =\prod_{a<b, \sigma (b)<\sigma (a)}
q_{i_{a}i_{b}}=q_{{\bf i},\sigma },$$
so the claim follows.

The proof of iii) is similar as for ii) with only difference that $\sigma $ is not unique.

\rem{\bf Remark 1.6.2}\ \  Note that for any weight $\nu =\sum \nu_{i}i$
with $|\nu |=
\sum \nu_{i}=n$, the size of the matrix $A^{(\nu )}$ is equal to the
multinomial coefficient $\frac{n!}{\prod_{i}\nu_{i}!}=\dim {\bf f}_{\nu }$,
in
particular for $\nu $ generic, $A^{(\nu )}$ is an $n!\times n!$ matrix.

\rem{\bf Example 1.6.3}\ \ Let $I=\{ 1,2,3\} $ and $\nu $ generic with
$\nu_{1}=\nu_{2}=\nu_{3}=1$. Then
w.r.t. basis $\{ \theta_{123}, \theta_{132}, \theta_{312}, \theta_{321},
\theta_{231}, \theta_{213}\} $
$$A^{123} = \left(\begin{array}{cccccc}
1 & q_{23} & q_{23}q_{13} & q_{12}q_{13}q_{23} & q_{12}q_{13} & q_{12} \\
q_{32} & 1 & q_{13} & q_{13}q_{12} & q_{12}q_{13}q_{32} & q_{12}q_{32} \\
q_{32}q_{31} & q_{31} & 1 & q_{12} & q_{12}q_{32} & q_{12}q_{31}q_{32} \\
\cdot & \cdot & \cdot & 1 & q_{32} & q_{31}q_{32} \\
\cdot & \cdot & \cdot & q_{23} & 1 & q_{31} \\
\cdot & \cdot & \cdot & q_{13}q_{23} & q_{13} & 1
\end{array} \right) =
\left(\begin{array}{cc}
X & Y \\
\bar{Y} & \bar{X}
\end{array} \right) $$
where $\bar{X}^{T}=X$, $Y^{T}=Y$.

\rem{\bf Example 1.6.4}\ \  Let $I=\{ 1,2,3\} $ and $\nu $ degenerate with
$\nu_{1}=2$, $\nu_{2}=0$, $\nu_{3}=1$. Then w.r.t. the basis
$\{ \theta_{113},\theta_{131},\theta_{311}\} $
$$A^{113} = \left(\begin{array}{ccc}
1+q_{11} & q_{13}+q_{11}q_{13} & q_{13}^{2}+q_{11}q_{13}^{2} \\
q_{31}+q_{31}q_{11} & 1+q_{11}q_{13}q_{31} & q_{13}+q_{11}q_{13} \\
q_{31}^{2}+q_{31}^{2}q_{11} & q_{31}+q_{31}q_{11} & 1+q_{11} \\
\end{array} \right). $$
Now we state some properties of the matrices $A^{(\nu )}$, $\nu $ generic,
which follow from the Proposition 1.6.1. For any sequences ${\bf i,j}$
of weight $\nu $ we have :

a)  $A^{(\nu )}_{{\bf i,i}} = 1$

b)  $A^{(\nu )}_{{\bf i,j}} = \overline{A^{(\nu )}_{{\bf j,i}}}$  ($A^{(\nu )}$ is
hermitian)

c)  $A^{(\nu )}_{{\bf \bar{i},\bar{j}}} = \overline{A^{(\nu )}_{{\bf i,j}}}$, where
${\bf \bar{i}}=i_{n}\dots i_{1}$ denotes the {\em reverse} of ${\bf i}=i_{1}
\dots i_{n}$

The property c) follows from the $\rho $-invariance 1.3.e) of $(\
,\ )_ {{\bf q}}$. Equivalently, we can write this in the matrix
form
$$P^{(\nu )}A^{(\nu )}P^{(\nu )}=\bar{A}^{(\nu )} (=(A^{(\nu )})^{T})$$
where $P^{(\nu )} (=(P^{(\nu )})^{-1})$ is the permutation matrix defined by $P_{{\bf i,j}}^{(\nu )}=\delta_{{\bf \bar{i},j}}$. As in our example for $n=3$,
one can also for general n write the matrix $A^{(\nu )}$, $\nu $ generic,
in the form
$\left(\begin{array}{cc}
X & Y \\
\bar{Y} & \bar{X}
\end{array} \right), $
with X hermitian and Y symmetric (e.g. if one uses the Johnson-Trotter
ordering of permutations (see [SWh],p.2).

\subsection{A reduction to generic case } 

Some questions about the matrices $A^{(\nu )}$ for general $\nu $ (e.g.
invertibility, positive definiteness) can be reduced to the generic situation
by using the following observation.

Let $\nu =\sum_{i}\nu_{i}i \in {\bf N}[I]$ be a degenerate weight. We shall
embed the matrix $A^{(\nu )}$ as a block in a block-diagonal
matrix associated to some generic weight. To do this let $\tilde{I}$ be
any set of size equal to $n=|\nu |=\sum_{i}\nu_{i}$ and let $\phi :
\tilde{I}\longrightarrow I$ be a function which maps exactly $\nu_{i}$
elements $ \tilde{i}$ of $\tilde{I}$ to $ i\in I$,
and let ${\bf \tilde{q}}$ be the induced hermitian family of parameters $
\tilde{q}_{\tilde{i},\tilde{j}}:= q_{i,j} (\tilde{i},\tilde{j}
\in \tilde{I})$ where $i=\phi ( \tilde{i}), j=\phi (\tilde{j})$.

Let ${\bf \tilde{f}}$ be the free associative algebra with generators
$\tilde
{\theta }_{1},\dots ,\tilde{\theta }_{n}$ and let $(\ ,\ )_{{\bf \tilde{q}}}$
be the sesquilinear form on ${\bf \tilde{f}}$ associated to ${\bf \tilde{q}}$
(as in 1.3). Let ${\bf \tilde{f}}_{\tilde{\nu }}$ be the generic weight space
corresponding to $\tilde{\nu }\in {\bf N}[\tilde{I}]$ where $\tilde{\nu }_{
\tilde{i}}=1$,
for every $\tilde{i}\in \tilde{I}$. Let $H=H_{\nu }$ be the group of all
bijections of $\tilde{I}$ which map $\phi^{-1}\{ i\} $ to itself for
every $i\in \phi(\tilde I)$. This group is isomorphic to the Young subgroup
$\prod_{i}S_{\nu_{i}} \subset S_{n}$. Let Y
be the subspace of ${\bf \tilde{f}}_{\tilde{\nu }}$ spanned by $H$-{\em
invariant vectors} $\tilde{\theta }_{H\tilde{{\bf i}}}=\sum_{h\in
H}\tilde{\theta }_{h\cdot \tilde{{\bf i}}}$ where
$\tilde{\theta }_{h\cdot \tilde{{\bf i}}} = \tilde\theta_{\tilde{ i}_
{h^{-1}(1)}}\cdots \tilde\theta_{\tilde{i}_{h^{-1}(n)}}$. Then for the
operator ${\bf \tilde{A}}$ associated to the form $(\ ,\ )_{\bf \tilde{q}}$
we have
$${\bf \tilde{A}}(\tilde{\theta }_{H\tilde{{\bf j}}})=\sum_{h\in H}
{\bf \tilde{A}}(\tilde{\theta }_{h\cdot {\bf \tilde{{\bf j}}}})=\sum_{h\in
H}\sum_{{\bf \tilde{i}}}
(\tilde{\theta }_{h\cdot {\bf \tilde{{\bf j}}}},\tilde{\theta }_{{\bf \tilde
{{\bf i}}}})_{{\bf \tilde{q}}}\tilde{\theta }_{{\bf \tilde{{\bf i}}}}.$$
Now for fixed $\tilde{{\bf i}}$ let $\tau $ be the unique permutation $\tau \in
S_{n}$ such that $\tilde{{\bf j}}=\tau \tilde{{\bf i}}$. So
\begin{eqnarray*}
\sum_{h\in H}(\tilde{\theta }_{h\cdot \tilde{{\bf j}}},\tilde{\theta }_
{\tilde{{\bf i}}})_{{\bf \tilde{q}}}&=&\sum_{h\in H}{\tilde{q}}_
{\tilde{{\bf i}},h\tau }\qquad \hbox{(by Prop.1.6.1.ii))}\\
&=&\sum_{h\in H}{q}_{{\bf i},h\tau }  {\hbox{ (by the definition of }}
{\tilde{q}}_{\tilde{{\bf i}},\tilde{{\bf j}}})=A^{(\nu )}_{{\bf i,j}}\
\hbox{ (by Prop. 1.6.1.iii)},
\end{eqnarray*}
where ${\bf i}=i_{1}\dots i_{n}=\phi (\tilde{i_{1}})\dots \phi
(\tilde{i_{n}})=:\phi ({\bf \tilde{i}})$ and
${\bf j}=\phi ({\bf \tilde{j}})$. Note that ${\bf j}=\phi (h{\bf
\tilde{j}})=\phi (h\tau {\bf \tilde{i}})=(h\tau )\cdot {\bf i}$, hence
${\bf \sigma } ({\bf i},{\bf j})=H\tau $. So we can write
$${\bf \tilde{A}}(\tilde{\theta }_{H{\bf \tilde{j}}})=\sum_{{\bf \tilde{i}}}
A_{{\bf i,j}}^{(\nu )}\tilde{\theta }_{{\bf \tilde{i}}}=\sum_{{\bf
i}} A_{{\bf i,j}}^{(\nu )}\tilde{\theta }_{H{\bf \tilde{i}}}$$
Thus we have proved that Y is an invariant subspace of the
operator ${\bf \tilde{A}}$ associated to the form $(\ ,\ )_{{\bf
\tilde{q}}}$ and moreover that the matrix of ${\bf \tilde{A}}|Y$
w.r.t the basis of $H$-invariant vectors $\tilde\theta_{H\bf
\tilde i}$ coincides with $A^{(\nu )}$. From this fact we conclude
that
\begin{enumerate}
\item[1)] If ${\bf \tilde{A}}|_{{\bf \tilde{f}}_{\tilde{\nu }}}$ is invertible, then
${\bf A}^{(\nu )}$ is invertible too. In particular
$$[A^{(\nu )}]^{-1}
_{{\bf i,j}}=\sum_{h\in H}[\tilde{A}^{(\tilde{\nu })}]^{-1}_{{\bf \tilde{i}},
h{\bf \tilde{j}}}$$ \\
where ${\bf \tilde{i},\tilde{j}}$ are chosen so that
$\phi ({\bf \tilde{i}})={\bf i}, \phi ({\bf \tilde{j}})={\bf j}$.
This means that the entries of $[A^{(\nu )}]^{-1}$, $\nu $ degenerate
can be read off from the sums of H-equivalent columns of the matrix
$[\tilde{A}^{(\tilde{\nu })}]^{-1}$, corresponding to the generic weight
$\tilde{\nu }$.
\item[2)] The determinant of $A^{(\nu )}$ divides the determinant
of $\tilde{A}^{(\tilde{\nu })}$.
\item[3)] If $\tilde{A}^{(\tilde{\nu })}$ is positive
definite, then $A^{(\nu )}$ is positive definite too.
\end{enumerate}

\subsection{Factorization of matrices $A^{(\nu )}$ for $\nu $ generic} 

First of all we point out that the rows of our multiparametric matrices
$A^{(\nu )}$ are not equal up to reordering (what was true in [Zag], where
all $q_{ij}$ are equal to q).

Therefore, the factorization of the matrices $A^{(\nu )}$ can not be reduced to
the factorization of the corresponding group algebra elements as was
treated by Zagier. Instead, by a somewhat tricky extension of the Zagier's
method  we show how this can be done on the matrix level\footnote{After
completing this paper it becomes clear that the matrix level computations can
be replaced by algebraic manipulations in a certain twisted group algebra and
then quasimultiplicative representations can be considered as ordinary
(multiplicative) representations of this twisted group algebra. This point of
view will be elaborated elsewhere.}.
This is achieved by
studying a $q_{ij}$-deformation of the regular representation  of the
symmetric group
which is only quasimultiplicative, i.e.,
multiplicative only up to factors which are diagonal ($q_{ij}$-dependent)
matrices
("projective representation").

Let $\nu =\sum \nu_{i}i \in {\bf N}[I]$ be a generic weight (i.e. $\nu_{i}\leq 1,
\forall i\in I$) and let $n=|\nu |=\sum \nu_{i}$. Let $R_{\nu }$ denote
the action of the symmetric group $S_{n}$ on the (generic) weight
space ${\bf f}_{\nu }$, given on the basis $B_{\nu }=\{ \theta_{{\bf i}}=
\theta_{i_{1}}\cdots \theta_{i_{n}}, |{\bf i}|=\nu \} $ of ${\bf f}_{\nu }$ by
place permutation,
$$R_{\nu }(g): \theta_{{\bf j}}=\theta_{j_{1}}\cdots \theta_{j_{n}}
\longrightarrow \theta_{g\cdot {\bf j}}=\theta_{j_{g^{-1}(1)}}\cdot
\theta_{j_{g^{-1}(n)}}.$$
Note that $g(k)$ indicates the place where the factor $\theta_{j_{k}}$
goes under the action $R_{\nu }(g)$.

Then $R_{\nu }$ is equivalent to the {\em right regular representation} $R_{n}$ of $
S_{n}$.

The corresponding {\em matrix representation}, also denoted by $R_{\nu }(g)$
is given by
$$R_{\nu }(g)_{{\bf i,j}}:=\delta_{{\bf i},g\cdot {\bf j}}.$$

Now, we need  more notations. Let $Q_{a,b}^{\nu }$ for
$1\leq a, b\leq n$ and
$Q^{\nu }(g)$, for $g\in S_{n}$ be the diagonal matrices
(multiplication operators on ${\bf f}_{\nu }$ ) defined by
$$(Q_{a,b}^{\nu })_{{\bf i,i}}:=q_{i_{a}i_{b}},$$
(e.g ($Q_{2,4}^{1234})_{4123,4123}=q_{13}$ if $I=\{ 1,2,3,4\},
\nu_{1}=\nu_{2}=\nu_{3}=\nu_{4} =1$)
$$Q^{\nu}(g)_{{\bf i,i}}:=q_{{\bf i},g^{-1}}=\prod_{a<b,
g^{-1}(a)>g^{-1}(b)}q_{i_{a}i_{b}} (\Longrightarrow
Q^{\nu}(g)=\prod_{(a,b)\in I(g^{-1})}Q^{\nu}_{a,b} ).$$
Note that $\bar{q}_{ij}=q_{ji}$ implies that $Q^{\nu}_{b,a}=[Q^{\nu}_{a,b}]^{*}$.
We also denote by $|Q^{\nu}_{a,b}|$ the diagonal matrix defined by
 $|Q^{\nu}_{a,b}|_{{\bf i,i}}=|q_{i_{a}i_{b}}|$. The quantity
$Q^{\nu}_{a,b}\cdot Q^{\nu}_{b,a}(=|Q^{\nu}_{a,b}|^{2})$ we abbreviate as
$Q^{\nu}_{\{ a,b\} }.$

More generally, for any subset $T\subseteq \{ 1,2,\cdots ,n\} $ we shall
use the notations
$$Q^{\nu}_{T}:=\prod_{a,b\in T, a\neq b}Q^{\nu}_{a,b}, \hbox{\large$\Box$}^{\nu}_{T}
:= I-Q^{\nu}_{T}$$
e.g. $Q^{\nu}_{\{ 3,5,6\} }=Q^{\nu}_{\{ 3,5\} }Q^{\nu}_{\{ 3,6\} }Q^{\nu}
_{\{ 5,6\} }=Q^{\nu}_{3,5}Q^{\nu}_{5,3}Q^{\nu}_{3,6}Q^{\nu}_{5,6}
Q^{\nu}_{6,5}$.

The following $q_{ij}$-{\em deformation of the right regular representation}
$R_{\nu }$, defined by
$$\hat{R}_{\nu }(g):=Q^{\nu}(g)R_{\nu }(g), \ g\in S_{n}$$
will be crucial in our method for factoring the matrices $A^{(\nu )}$
$\nu $-generic.
\bthm{PROPOSITION 1.8.1.} If $\nu $ is a generic weight with $|\nu |=n$, then
for the matrix $A^{(\nu )}$ of $(\ ,\ )_{\bf q}$ on ${\bf f}_\nu$ we have
$$A^{(\nu )}=\sum_{g\in S_{n}}\hat{R}_{\nu }(g)$$
\ethm
{\bf Proof.}  The (${\bf i,j}$)-th entry of the r.h.s. is equal to \\
$\sum_{g\in S_{n}}\hat{R}_{\nu }(g)_{{\bf i,j}}=
\sum_{g\in S_{n}}Q(g)_{{\bf i,i}}{R}_{\nu }(g)_{{\bf i,j}}=
\sum_{g\in S_{n}}q_{{\bf i},g^{-1}}\delta_{{\bf i},g\cdot {\bf j}}=
q_{{\bf i},\tau ^{-1}}$, if ${\bf i}=\tau {\bf j}$ (such $\tau $ is unique,
because $|{\bf i}|=|{\bf j}|=\nu $ is generic), what is just $A^{(\nu )}
_{{\bf i,j}}$, according to Prop.1.6.1 ii) and the proof follows.\vskip10pt

Before we proceed with the factorization of matrices $A^{(\nu )}$ we need
more detailed information concerning our "projective" right regular
representation $\hat{R}_{\nu }$ which is only
quasimultiplicative in the following sense:\vskip10pt\noindent
{PROPERTY 0.} (quasimultiplicativity)
$$\hat{R}_{\nu }(g_{1})\hat{R}_{\nu }(g_{2})=\hat{R}_{\nu }(g_{1}g_{2})
\hbox{ \ if \ } l(g_{1}g_{2})=l(g_{1})+l(g_{2})$$
where $l(g):=Card\ I(g)$ is the length of $g\in S_{n}$.\vskip10pt

This property follows from the following general formula :
\bthm{PROPOSITION 1.8.2.} For any $g_{1},g_{2}\in S_{n}$ we have
$$\hat{R}_{\nu }(g_{1})\hat{R}_{\nu }(g_{2})=M_{\nu }(g_{1},g_{2})
\hat{R}_{\nu }(g_{1}g_{2})$$
where the multiplication factor is the diagonal matrix
$$M_{\nu }(g_{1},g_{2})=\prod_{(a,b)\in I(g_{1}^{-1})-I(g_{2}^{-1}g_{1}^{-1})}
Q^{\nu}_{\{ a,b\} }\quad(=\prod_{(a,b)\in I(g_{1})\cap I(g_{2}^{-1})}Q^{\nu}_
{\{ g_{1}(a),
g_{1}(b)\} }).$$
\ethm
{\bf Proof.}  First we observe that for any diagonal matrix D, its conjugate
by the "permutation" matrix $R(g)$, $D^{(g)}=R(g)DR(g)^{-1}$ is a diagonal matrix such that $D^{(g)}_{{\bf i,i}}=D_{g^{-1}\cdot {\bf i},g^{-1}\cdot {\bf i}}$.
Then by the definition of $\hat{R}_{\nu }$ and writing $Q$ instead
of $Q^{\nu}$ we obtain:
\begin{eqnarray*}
\hat{R}_{\nu }(g_{1})\hat{R}_{\nu }(g_{2})&=&Q(g_{1})[R_{\nu }(g_{1})Q(g_{2})]
R_{\nu }(g_{2})=Q(g_{1})Q(g_{2})^{(g_{1})}R_{\nu }(g_{1})R_{\nu }(g_{2})\\
&=&Q(g_{1})Q(g_{2})^{(g_{1})}R_{\nu }(g_{1}g_{2})=Q(g_{1})Q(g_{2})^{(g_{1})}
Q(g_{1}g_{2})^{-1}\hat{R}_{\nu }(g_{1}g_{2})
\end{eqnarray*}
i.e. $M_{\nu }(g_{1},g_{2})=Q(g_{1})Q(g_{2})^{(g_{1})}Q(g_{1}g_{2})^{-1}$.

By using that $[Q_{a,b}^{(g)}]_{{\bf i,i}}=[Q_{a,b}]_{g^{-1}\cdot {\bf i},
g^{-1}\cdot {\bf i}}=q_{i_{g(a)}i_{g(b)}} (\Longrightarrow Q_{a,b}^{(g)}
=Q_{g(a),g(b)})$ we can rewrite and split $Q(g_{2})^{(g_{1})}$ and $Q(g_{1}
g_{2})$ as follows :
\begin{eqnarray*}
Q(g_{2})^{(g_{1})}&=&\left[\prod_{(a^{'},b^{'})\in I(g_{2}^{-1})}Q_{a^{'},b^{'}}\right]^{(
g_{1})}=\prod_{(a^{'}, b^{'} )\in I(g_{2}^{-1})}Q_{a^{'},b^{'}}^{(g_{1})}
=\prod_{(a^{'},b^{'} )\in I(g_{2}^{-1})}Q_{g_{1}(a^{'})g_{1}(b^{'})} \\
&=&\prod_{(g_{1}^{-1}(a),g_{1}^{-1}(b))\in I(g_{2}^{-1})}Q_{a,b}
=\prod_{(a,b)\in I(g_{2}^{-1}g_{1}^{-1})-I(g_{1}^{-1})}Q_{a,b}\cdot
\prod_{(b,a)\in I(g_{1}^{-1})-I(g_{2}^{-1}g_{1}^{-1})}Q_{a,b}\\[0pt\pagebreak]
Q(g_{1}g_{2})&=&\prod_{(a,b)\in I(g_{2}^{-1}g_{1}^{-1})}Q_{a,b} \\
&=&\prod_{(a,b)\in I(g_{1}^{-1})\cap I(g_{2}^{-1}g_{1}^{-1})}Q_{a,b}\cdot
\prod_{(a,b)\in I(g_{2}^{-1}g_{1}^{-1})-I(g_{1}^{-1})}Q_{a,b}=Q^{'}\cdot
Q^{''}
\end{eqnarray*}
Finally, since diagonal matrices commute, after cancellation, we get
\begin{eqnarray*}
M_{\nu}(g_{1},g_{2})&=&[Q(g_{1})Q^{{'}^{-1}}][Q(g_{2})^{(g_{1})}Q^{{''}^{-1}}]\\
&=&\prod_{(a,b)\in I(g_{1}^{-1})-I(g_{2}^{-1}g_{1}^{-1})}Q_{a,b}\prod_{(b,a)\in
I(g_{1}^{-1})-I(g_{2}^{-1}g_{1}^{-1})}Q_{a,b}\\
&=&\prod_{(a,b)\in I(g_{1}^{-1})-I(g_{2}^{-1}g_{1}^{-1})}Q_{\{ a,b\} },
\end{eqnarray*}
and the proof is finished.\vskip10pt

For $1\leq a\leq b\leq n$ we denote by $t_{a,b}$ the following cyclic
permutation in $S_{n}$
$$t_{a,b}:=\left( \begin{array}{cccc}
a & a+1 & \cdots & b \\
b & a & \cdots & b-1
\end{array} \right) $$
which maps $b$ to $b-1$ to $b-2$ $\cdots $ to $a$ to $b$ and fixes all $1\leq k
<a$ and $b<k\leq n$. Its inverse is then
$$t_{a,b}^{-1}=\left( \begin{array}{cccc}
a & a+1 & \cdots & b \\
a+1 & a+2 & \cdots & a
\end{array} \right). $$
Note that the corresponding sets of inversions are equal to $I(t_{a,b})=
\{ (a,j) | a<j\leq b\} $ and $I(t_{a,b}^{-1})=\{ (i,b) | a\leq i<b\} $.

We also denote by
$$t_{a}:=t_{a,a+1} (1\leq a<n)$$
the transposition of adjacent letters $a$ and $a+1$.

Then, from Proposition 1.8.2, one gets the following more specific properties
of $\hat{R}_{\nu }$ which we shall need later on:\vskip10pt\noindent
PROPERTY 1. (braid relations)
\begin{eqnarray*}
&&\hat{R}_{\nu }(t_{a})\hat{R}_{\nu }(t_{a+1})\hat{R}_{\nu }(t_{a})=
\hat{R}_{\nu }(t_{a+1})\hat{R}_{\nu }(t_{a})\hat{R}_{\nu }(t_{a+1}),
\forall a=1,\dots ,n-2\\
&&\hat{R}_{\nu }(t_{a})\hat{R}_{\nu }(t_{b})=\hat{R}_{\nu }(t_{b})\hat{R}_{\nu
}(t_{a}), \forall a,b=1,\dots ,n-1 \hbox{ with } |a-b|\geq 2.
\end{eqnarray*}
PROPERTY 2.
$$\hat{R}_{\nu }(g)\hat{R}_{\nu }(t_{a,b})=\prod_{a\leq i<b, g(i)>g(b)}
Q^{\nu}_{\{ g(b),g(i)\} }\hat{R}_{\nu }(gt_{a,b}),$$
for $g\in S_{n},\ \ 1\leq a<b\leq n.$
In particular we have\vskip10pt\noindent
PROPERTY $2^{'}$.
$$\hat{R}_{\nu }(g)\hat{R}_{\nu }(t_{k,m})=\hat{R}_{\nu }(gt_{k,m}),$$
for $g\in S_{m-1}\times S_{n-m+1}, 1\leq k\leq m\leq n$.\vskip10pt\noindent
PROPERTY 3. (commutation rules) i) For $1\leq a\leq a^{'}<m\leq n$
$$\hat{R}_{\nu }(t_{a^{'},m})\hat{R}_{\nu }(t_{a,m})=Q^{\nu}_{\{ m-1,m\} }\hat{R}
_{\nu }(t_{a,m-1})\hat{R}_{\nu }(t_{a^{'}+1,m}).$$
ii) Let $w_{n}=n \: n-1\cdots 2 \: 1$ be the longest permutation in $S_{n}$.
Then for any $g\in S_{n}$
\begin{eqnarray*}
\hat{R}_{\nu}(gw_{n})\hat{R}_{\nu}(w_{n})&=&\hat{R}_{\nu}(w_{n})\hat{R}
_{\nu}(w_{n}g)\\
&=&(\prod_{a<b, g^{-1}(a)<g^{-1}(b)}Q^{\nu}_{\{ a,b\}})\hat{R}(g)\
(=|Q^{\nu}(gw_{n})|^{2}\hat{R}(g))
\end{eqnarray*}
PROPERTY 4.  For any $1\leq a_{1}<a_{2}<\cdots <a_{s}<m\leq n$, we have
$$\hat{R}_{\nu }(t_{a_{1},m})\hat{R}_{\nu }(t_{a_{2},m})\cdots \hat{R}_{\nu }
(t_{a_{s},m})=\hat{R}_{\nu }(t_{a_{1},m}t_{a_{2},m}\cdots t_{a_{s},m}).$$
Now we can state our first factorization of the matrices $A^{(\nu )}$, $\nu $
generic.
\bthm{PROPOSITION 1.8.3.} For $1\leq m\leq n$, we define
$$A^{(\nu ),m}:=\hat{R}_{\nu }(t_{1,m})+\hat{R}_{\nu }(t_{2,m})+\cdots +\hat{R}_
{\nu }(t_{m,m})\ \  (A^{(\nu ),1}=I).$$
Then we have the following factorization
$$A^{(\nu )}=A^{(\nu ),1}A^{(\nu ),2}\cdots A^{(\nu ),n}.$$
\ethm
{\bf Proof.}  Since any element $g\in S_{n}$ can be represented uniquely
as $g_{1}t_{k,n}$, with $g_{1}\in S_{n-1}\times S_{1}\subset
S_{n}$ and $1\leq k\leq n$ (namely $k=g^{-1}(n), g_{1}=gt^{-1}_{k,n}$), we
can write
\begin{eqnarray*}
A^{(\nu )}&=&\sum_{g\in S_{n}}\hat{R}_{\nu }(g)=\sum_{g_{1}\in
S_{n-1}\times S_{1}, 1\leq k\leq n}\hat{R}_{\nu }(g_{1}t_{k,n})\\
&=& (\sum_{g_{1}\in S_{n-1}\times S_{1}}\hat{R}_{\nu }(g_{1}))
(\sum_{k=1}^{n}\hat{R}_{\nu }(t_{k,n}))
\end{eqnarray*}
where the first equality is by Prop.1.8.1 and the third equality follows by
Property $2^{'}$. Subsequently, we represent $g_{1}\in
S_{n-1}\times S_{1}$ uniquely as $g_{1}=g_{2}t_{k_{2},n-1}$
with $g_{2}\in S_{n-1}\times S_{1}^{2}$ and $1\leq k_{2}
\leq n-1$ and so on. The claim follows.

We now make a second reduction by expressing the matrices $A^{(\nu ),m}$ in
turn as products of yet simpler matrices.
\bthm{PROPOSITION 1.8.4.} Let $C^{(\nu ),m} (m\leq n)$ and $D^{(\nu ),m}
(m<n)$ be the following matrices\newline
$C^{(\nu ),m}:=[I-\hat{R}_{\nu }(t_{1,m})][I-\hat{R}_{\nu }(t_{2,m})]\cdots
[I-\hat{R}_{\nu }(t_{m-1,m})]$,\newline
$D^{(\nu ),m}=[I-Q^{\nu}_{\{ m,m+1\} }\hat{R}_{\nu }(t_{1,m})]
[I-Q^{\nu}_{\{ m,m+1\} }\hat{R}_{\nu }(t_{2,m})]\cdots
[I-Q^{\nu}_{\{ m,m+1\} }\hat{R}_{\nu }(t_{m,m})]$.\newline
Then
$$A^{(\nu ),m}=D^{(\nu ),m-1}[C^{(\nu ),m}]^{-1}$$
\ethm\vskip-\lastskip
{\bf Proof.}\  Let $A^{(\nu ),r,m}:=\sum_{k=r}^{m}\hat{R}(t_{k,m})$, so that
$A^{(\nu ),1,m}=A^{(\nu ),m}, A^{(\nu ),m,m}=I$ (because $t_{m,m}=1\in
S_{n}$). By using Property 3. (commutation rules) we find
\begin{eqnarray*}
A^{(\nu ),r,m}(I-\hat{R}_{\nu}(t_{r,m}))&=&\hat{R}_{\nu}(t_{r,m})+
\sum_{k=r+1}^{m}
\hat{R}_{\nu}(t_{k,m})-\sum_{k=r}^{m-1}\hat{R}_{\nu}(t_{k,m})
\hat{R}_{\nu}(t_{r,m})-\hat{R}_{\nu}
(t_{r,m})\\
&=&\sum_{k=r+1}^{m}\hat{R}_{\nu}(t_{k,m})-\sum_{k=r+1}^{m}Q^{\nu}_{\{ m-1,m\}
} \hat{R}_{\nu}(t_{r,m-1}) \hat{R}_{\nu}(t_{k,m})\\
&=&(I-Q^{\nu}_{\{ m-1,m\} }\hat{R}(t_{r,m-1}))A^{(\nu ),r+1,m}
\end{eqnarray*}
and hence by induction on $r$ (starting with the trivial case $r=0$)
\begin{eqnarray*}
&&A^{(\nu ),1,m}[I-\hat{R}_{\nu}(t_{1,m})]\cdots [I-\hat{R}_{\nu}(t_{r,m})]=\\
&&=[I-Q^{\nu}_{\{ m-1,m\} }\hat{R}_{\nu}(t_{1,m-1})]\cdots
[I-Q^{\nu}_{\{ m-1,m\} }\hat{R}_{\nu}(t_{r,m-1})
A^{(\nu ),r+1,m}].
\end{eqnarray*}
The case $r=m-1$ of this identity is the desired identity.

\subsection{Formula for the determinant of $A^{(\nu )}$, $\nu $ generic.} 

So far we have expressed the matrix $A^{(\nu )}$ as a product of
matrices like $I-Q^{\nu}_{\{ m,m+1\} }\hat{R}_{\nu}(t_{k,m})$ or
$[I-\hat{R}_{\nu}(t_{k,m})]^{-1}$.
Thus, in order to evaluate $\det A^{(\nu )}$, we first compute the determinant
of such matrices.
\bthm{LEMMA 1.9.1.} For $\nu $ generic with $|\nu |=n$, we have\newline
a) $\dsty\det(I-\hat{R}_{\nu }(t_{a,b}))=\prod_{\mu\subseteq \nu , |\mu|=b-a+1}
(\Box_{\mu})^{(b-a)!(n+a-b-1)!} , (1\leq a<b\leq n)$\newline
b) $\dsty\det(I-Q^{\nu}_{\{ b,b+1\} }\hat{R}_{\nu}(t_{a,b}))=
\prod_{\mu\subseteq \nu , |\mu|=b-a+2}(\Box_{\mu})
^{(b-a)!(b-a+2)!(n+a-b-2)!}, (1\leq a\leq b<n)$\newline
where for any subset $T\subset I$ we denote by $\Box_{T}$
the quantity
$$\Box_{T}:=1-{q}_{T} ; \ \ {q}_{T}=\prod_{i\neq j\in T}q_{ij}
(=\prod_{\{ i\neq j\} \subset T}|q_{ij}|^{2})$$
in which the last product is over all two-element subsets
of $T$ (We view $\nu $ as a subset of $I$, hence $\mu\subseteq \nu $
means that $\mu$ is a subset of $\nu$).
\ethm
{\bf Proof.} a) Let $H:=<t_{a,b}>\subset S_{n}$ be the cyclic
subgroup of $S_{n}$ generated by the cycle $t_{a,b}$.
Then, each $H$-orbit on ${\bf f}_{\nu }$, ${\bf f}_{\nu }^{[{\bf i}]_
{a}^{b}}=span\{
\theta_{t_{a,b}^{k}\cdot {\bf i}} | 0\leq k\leq b-a\} $, (which clearly
corresponds to a cyclic $t_{a,b}$-equivalence class $[{\bf i}]^{b}_{a}
=i_{1}\cdots (i_{a}i_{a+1}\cdots i_{b})\cdots i_{n}$ of the sequence
${\bf i}=i_{1}\dots i_{n}$ of weight $\nu $) is an invariant subspace of
$R_{\nu }(t_{a,b})$ (and hence of $\hat{R}_{\nu }(t_{a,b})$). Note that $\hat{R}
_{\nu }(t_{a,b})(\theta_{t_{a,b}^{k}\cdot {\bf i}})=c_{k}\theta_{t_{a,b}^{k+1}
\cdot {\bf i}}$ where $c_{k}=q_{t_{a,b}^{k}\cdot {\bf i},
t_{a,b}^{-1}} (0\leq k\leq b-a)$ i.e.
\begin{eqnarray*}
c_{0}&=&q_{i_{a}i_{b}}q_{i_{a+1}i_{b}}\cdots q_{i_{b-1}i_{b}},\\
c_{1}&=&q_{i_{a+1}i_{a}}q_{i_{a+2}i_{a}}\cdots q_{i_{b}i_{a}},\\
&\vdots&\\
c_{b-a}&=&q_{i_{b}i_{b-1}}q_{i_{a}i_{b-1}}\cdots q_{i_{b-2}i_{b-1}},
\end{eqnarray*}
i.e. $\hat{R}_{\nu }(t_{a,b})|f_{\nu }^{[{\bf i}]_{a}^{b}}$ is a
cyclic operator, hence
\begin{eqnarray*}
\det(I-\hat{R}_{\nu }(t_{a,b})|f_{\nu }^{[{\bf i}]_{a}^{b}})&=&1-c_{0}c_{1}
\cdots c_{b-a}=1-\prod_{i\neq j\in \{ i_{a},\dots ,i_{b}\} }q_{ij}\\
&=&1-\prod_{\{ i,j\} \subset \{ i_{a},\dots ,i_{b}\} }|q_{ij}|
^{2}=\Box_{\{ i_{a},\dots ,i_{b}\} }.
\end{eqnarray*}
Note that this determinant depends only on the set $ \{ i_{a},i_{a+1},
\dots, i_{b}\} $ and that there are $(b-a)!(n-(b-a+1))!$ cyclic
$t_{a,b}$ -equivalence classes corresponding to any given
$(b-a+1)$-set $\mu=\{ i_{a},\dots ,i_{b}\} \subset \nu $.
(Here we identify a generic weight $\nu =\sum \nu_{i}\cdot
i$, $\nu_{i}\leq 1$ with the set $\{ i\in I|\nu_{i}=1\} $).

b)  Similary as in a) we have $Q^{\nu}_{\{ b,b+1\} }\hat{R}_{\nu }(t_{a,b})
(\theta_{t_{a,b}^{k}
\cdot {\bf i}})=d_{k}\theta_{t_{a,b}^{k+1}\cdot {\bf i}} (0\leq k\leq b-a)$
where $d_{0}=c_{0}|q_{i_{b}i_{b+1}}|^{2},
d_{1}=c_{1}|q_{i_{a}i_{b+1}}|^{2}$, $\dots $,
$d_{b-a}=c_{b-a}|q_{i_{b-1}i_{b+1}}|^{2}$ (with $c_{k}$ as above).
Then \newline
$\det(I-Q^{\nu}_{\{ b,b+1\} }\hat{R}_{\nu }(t_{a,b})|f_{\nu }^{[{\bf i}]_
{a}^{b}})=
1-d_{0}d_{1}\cdots d_{b-a}=1-\prod_{\{ i,j\} \subset \{ i_{a},\dots ,
i_{b+1}\} }|q_{ij}|^{2}=\Box_{\{ i_{a},\dots ,i_{b+1}\} }$.

Now, for given $(b-a+2)$-set $\mu\subset \nu $, we shall count the number of $H$-orbits labeled by $[{\bf i}]_{a}^{b}$ on which the above determinant
assumes the same value $\Box_{\mu}$.
We can choose any element of $\mu$ to be $i_{b+1}$ (in $b-a+2$ ways), then
the remaining $b-a+1$ elements in $\mu$ can be arranged in  $(b-a+1)!/(b-a+1)
=(b-a)!$ cyclic arrangements $(i_{a}\cdots i_{b})$ and the remaining $n-(b-a+2)
$ positions in $[{\bf i}]^{b}_{a}=i_{1}\cdots (i_{a}\cdots i_{b})
i_{b+1}\cdots i_{n}$ can form any permutation of the set $\nu -\mu$ (in
$(n+a-b-2)!$ ways).
\bthm{THEOREM 1.9.2. [THE DETERMINANTAL FORMULA]} The determinant of the matrix $A^{(\nu )}$, $\nu $ generic,
is given by
$$\det A^{(\nu )}=\prod_{\mu\subseteq \nu, |\mu| \geq 2 }(\Box_{\mu})^{(|\mu|-2)!(|\nu
|-|\mu|+1)!}.$$
\ethm\vskip-\lastskip
{\bf Proof.}  By Lemma 1.9.1 applied to matrices $C^{(\nu ),m}, D^{(\nu ),
m-1}$ (defined in Prop.1.8.4), we have
\begin{eqnarray*}
\det C^{(\nu ),m}&=&\prod_{\mu\subseteq \nu , 2\leq |\mu|\leq m}
(\Box_{\mu})^{(|\mu|-1)!(n-|\mu|)!}\\
\det D^{(\nu ),m-1}&=&\prod_{\mu\subseteq \nu , 2\leq |\mu|\leq m}
(\Box_{\mu})^{(|\mu|-2)!|\mu|(n-|\mu|)!}
\end{eqnarray*}
Then, by Prop 1.8.4
$$\det A^{(\nu ),m}=\det D^{(\nu ),m-1}/\det C^{(\nu ),m}=\prod_{\mu\subseteq
\nu , 2\leq |\mu|\leq m}(\Box_{\mu})^{(|\mu|-2)!(n-|\mu|)!} $$
Finally, by Prop 1.8.3
$$\det A^{(\nu )}=\prod_{m=1}^{n}\det A^{(\nu ),m}=\prod_{\mu\subseteq
\nu, |\mu| \geq 2 }(\Box_{\mu})^{(|\mu|-2)!(n-|\mu|+1)!}.$$
This completes the proof.

In particular, in Example 1.6.3 ($I=\{ 1,2,3\} ,\nu_{1}=\nu_{2}=\nu_{3}=1$)
we have
$$\det A^{123}=(1-|q_{12}|^{2})^{2}(1-|q_{13}|^{2})^{2}
(1-|q_{23}|^{2})^{2}(1-|q_{12}|^{2}|q_{13}|^{2}|q_{23}|^{2})$$
{\bf Remark 1.9.3.}  Theorem 1.9.2 represents a multiparametric extension
of the Theorem 2 in [Zag] which states that in one-parametric case
$(q_{ij}=q)$:
$$\det A_{n}(q)=\prod_{k=2}^{n}(1-q^{k(k-1)})^{\frac{n!(n-k+1)}{k(k-1)}}$$
(e.g. $\det A_{3}(q)=(1-q^{2})^{6}(1-q^{6})$).
\bthm{THEOREM 1.9.4.} The matrix $A=A({\bf q})$ associated to the sesquilinear
form $(\ , \ )=(\ ,\ )_{{\bf q}}$ on ${\bf f}$, (see 1.3 and 1.6) is
positive definite if $|q_{ij}|<1$, for all $i,j\in I$, so that the
$q_{ij}$-cannonical commutation relations 1.1(1) have a Hilbert space
realization (cf. 1.5).
\ethm
{\bf Proof.}  From Prop.1.6.1 we know that $A=\bigoplus A^{(\nu )}$.
For $\nu $ generic, we see directly from Theorem 1.9.2 that $A^{(\nu )}$
is nonsingular if $|q_{ij}|<1$ for all $ i\neq j\in I$.
According to the reduction to the generic case (discussed in 1.7) we
see that $A^{(\nu )}$, $\nu $ degenerate, is also nonsingular if $|q_{
ij}|<1$, for all $i,j\in I$. Since $A({\bf 0})$ (i.e. if all
$q_{ij}=0$) is the identity matrix and the eigenvalues of $A({\bf q})$
vary continuously with $q_{ij}$ and are real (because $A({\bf q})$ is
hermitian) we see that $A({\bf q})$ is positive definite if $|q_{ij}|<1,
i,j\in I$.

\section{Formulas for the inverse of $A^{(\nu )}$, $\nu $ generic.} 

The problem of computing the inverse of matrices $A^{(\nu )}$
appears in the expansions of the number operators and transition
operators (c.f [MSP]). It is also related to a random walk problem
on symmetric groups and in several other situations (hyperplane
arrangements, contravariant forms on certain quantum groups).
We shall give here two types of formulas for $[A^{(\nu )}]^{-1}$: a Zagier
type formula and Bo\v zejko-Speicher type formulas.

\subsection{Zagier type formula} 

First we give a formula for the inverse of $A^{(\nu )}$, $\nu $ generic,
which follows from Prop.1.8.3 and Prop.1.8.4 :
\begin{eqnarray*}
[A^{(\nu )}]^{-1}&=&[A^{(\nu ),n}]^{-1}\cdots [A^{(\nu ),1}]^{-1}\\
&=&C^{(\nu ),n}\cdot [D^{(\nu ),n-1}]^{-1}\cdot C^{(\nu ),n-1}\cdot [D^{(\nu
),n-2}]^{-1}\cdots C^{(\nu ),2}\cdot [D^{(\nu ),1}]^{-1}
\end{eqnarray*}
To invert $A^{(\nu )}$, therefore, the first step is to invert $D^{(\nu )
,m}$ for each $m<n$. First we recall the notation
$Q_{T}^\nu=\prod_{a,b\in T,a\neq b} Q_{a,b}^\nu$, $\bbox_T^\nu=I-Q_T^\nu$
$(T\subseteq\{1,2,\dots,n\})$ from 1.8.

\bthm{PROPOSITION 2.1.1.} For $\pi \in S_{n}$ let $Des(\pi )$ denote the
descent set of $\pi $ (i.e. the set $\{ 1\leq i\leq n-1 | \pi (i
)>\pi (i+1)\} $) and let $W^{\nu}_{m}(\pi ) (m<n)$ be the following
diagonal matrix
$$W^{\nu}_{m}(\pi )=\prod_{i\in Des(\pi^{-1})}Q^{\nu}_{[i+1..m+1]}.$$
Then the inverse of the matrix $D^{(\nu ),m}$ is given explicitly by
$$[D^{(\nu ),m}]^{-1}=[\bigtriangleup ^{(\nu ),m}]^{-1}E^{(\nu ),m}$$
where
$$E^{(\nu ),m}=\sum_{\pi \in S_{m}\times S_{1}^{n-m}}
W^{\nu}_{m}(\pi )\hat{R}_{\nu }(\pi )$$
and where $\bigtriangleup ^{(\nu ),m}$ is the following diagonal matrix
$$\bigtriangleup ^{(\nu ),m}:=\bbox^{\nu}_{[1..m+1]}\bbox^{\nu}_{[2..m+1]}
\cdots \bbox^{\nu}_{[m..m+1]}$$
(Here $[a..b]$ denotes the set $\{ a,a+1,\cdots ,b\} $).
\ethm
{\bf Proof.}  Denote by ${\bf \sigma }\longrightarrow {\bf \tilde
{\sigma }}$ the obvious map $S_{m}\times S_{1}^{n-m}
\longrightarrow S_{1}\times S_{m}\times S_{1}
^{n-m-1}$ (i.e. $ \tilde{\sigma }(1)=1, \tilde{\sigma }(i)=
 \sigma (i-1)+1$ for $i>1$). This is a homomorphism since
$ \tilde{\sigma }=t_{1,n}^{-1}\sigma t_{1,n}$ (because $m<n$).
It is easy to check that then
$$\hat{R}({\bf \tilde{\sigma }})=R(t_{1,n})^{-1}\hat{R}({\bf \sigma })
R(t_{1,n})$$
Also we note that $\tilde{t_{a,b}}=t_{a+1,b+1}$, for $1\leq a<b\leq m$.
Thus we can rewrite the matrix $D^{(\nu ),m}$ as follows:
$$D^{(\nu ),m}=[I-Q^{\nu}_{\{ m,m+1\} }\hat{R}_{\nu}(t_{1,m})]\tilde{D}^
{(\nu ),m-1}$$
where we set
\begin{eqnarray*}
\tilde{D}^{(\nu ),m-1}&:=&[I-Q^{\nu}_{\{ m,m+1\} }\hat{R}_{\nu}(t_{2,m})]
\cdots [I-Q^{\nu}_{\{ m,m+1\} }
\hat{R}_{\nu}(t_{m,m})]\\
&=&[I-Q^{\nu}_{\{ m,m+1\} }\hat{R}_{\nu}(\tilde{t}_{1,m-1})]\cdots
[I-Q^{\nu}_{\{ m,m+1\} }\hat{R}_{\nu}(\tilde{t}_{m-1,m-1})]
\end{eqnarray*}
By noting that $R_{\nu}(t_{1,n})Q^{\nu}_{\{ m,m+1\} }=Q^{\nu}_{\{ m-1,m\} }
R_{\nu}(t_{1,n})$\newline
($\Longrightarrow Q^{\nu}_{\{ m,m+1\} }=R_{\nu}(t_{1,n})^{-1}
Q^{\nu}_{\{ m-1,m\} }R_{\nu}(t_{1,n})$) we have
\begin{eqnarray*}
\tilde{D}^{(\nu ),m-1}&=&R_{\nu}(t_{1,n})^{-1}[I-Q^{\nu}_{\{ m-1,m\} }
\hat{R}_{\nu}(t_{1,m-1})]
\cdots [I-Q^{\nu}_{\{ m-1,m\} }\hat{R}_{\nu}(t_{m-1,m-1})]R_{\nu}(t_{1,n})\\
&=&R_{\nu}(t_{1,n})^{-1}D^{(\nu ),m-1}R_{\nu}(t_{1,n})
\end{eqnarray*}
Therefore, to prove the formula $[D^{(\nu ),m}]^{-1}=[\bigtriangleup^{(\nu
),m}]^{-1}E^{(\nu ),m}$ by induction, it suffices to show that
$$E^{(\nu ),m}(I-Q^{\nu}_{\{ m,m+1\} }\hat{R}_{\nu}(t_{1,m}))=
[1-Q^{\nu}_{[1..m+1]}]\tilde{E}^{(\nu ),m-1} \eqno(*)$$
where we set
$$\tilde{E}^{(\nu ),m-1}=R_{\nu}(t_{1,n})^{-1}E^{(\nu ),m-1}R_{\nu}(t_{1,n})$$
To show (*), we first calculate
\begin{eqnarray*}
&&E^{(\nu ),m}Q^{\nu}_{\{ m,m+1\} }\hat{R}_{\nu}(t_{1,m})=\sum_{\sigma \in
S_{m} \times S_{1}^{n-m}}W^{\nu}_{m}(\sigma )\hat{R}_{\nu}(\sigma )
Q^{\nu}_{\{ m,m+1\} }\hat{R}_{\nu}(t_{1,m})\\
&&=\sum_{\sigma \in S_{m}\times S_{1}^{n-m}}W^{\nu}_{m}(\sigma )
Q^{\nu}_{\{ \sigma (m),\sigma (m+1)\} }\hat{R}_{\nu}(\sigma )
\hat{R}_{\nu}(t_{1,m})\\
&&\hbox to 0pt{By using that $\sigma (m+1)=m+1$, and by Property 2 (stated in 1.8)\hss}\\
&&=\sum_{\sigma \in S_{m}\times S_{1}^{n-m}}W^{\nu}_{m}(\sigma )
Q^{\nu}_{\{ \sigma (m),m+1\} }\prod_{m\geq j>\sigma (m)}Q^{\nu}_{\{ \sigma (m),
j\} }\hat{R}_{\nu}(\sigma t_{1,m})\\
&&=\sum_{\pi \in S_{m}\times S_{1}^{n-m}}W^{\nu}_{m}(\pi t_{1,m}^{-1})
\prod_{\pi (1)<j\leq m+1}Q^{\nu}_{\{ \pi (1),j\} }\hat{R}_{\nu}(\pi )
\end{eqnarray*}
By observing that the descent sets of $\pi^{-1}$ and $(\pi t_{1,m}^{-1})
^{-1}=t_{1,m}\pi^{-1}$ are related by
$$Des(t_{1,m}\pi^{-1})=\cases{(Des(\pi^{-1})\setminus\{ \pi (1)-1\} )\bigcup
\{ \pi (1)\}, & if $\pi (1)>1$\cr
Des(\pi^{-1})\bigcup \{ \pi (1)\}, & if $\pi (1)=1$\cr}$$
we see immediately that
$$W^{\nu}_{m}(\pi t_{1,m}^{-1})\cdot \prod_{\pi (1)<j\leq m+1}
Q^{\nu}_{\{ \pi (1),j\} }=\cases{W^{\nu}_{m}(\pi ), & if $\pi (1)>1$ \cr
Q^{\nu}_{[1..m+1]}W^{\nu}_{m}(\pi ), & if $\pi (1)=1$\cr}$$
By plugging this into the l.h.s. of (*), after cancellation, we obtain
the r.h.s. of (*). This completes the proof of Prop.2.1.1.

We next give a formula expressing the matrices $C^{(\nu ),m} (m\leq n)$
as a sum rather than a product (see Prop.1.8.4).
\bthm{PROPOSITION 2.1.2.} The matrices $C^{(\nu ),m}, (m\leq n)$ defined in
Prop.1.8.4 are given by
$$ C^{(\nu ),m}=\sum_{k=1}^{n}C^{(\nu ),m;k}, \ C^{(\nu ),m;k}=
(-1)^{m-k}\sum_{\pi \in S_{m}^{(k)}\times S_{1}^{n-m}}
\hat{R}_{\nu}(\pi^{-1})$$
where $S_{m}^{(k)}$ is the subset of $S_{m}$ of cardinality ${m-1 \choose
k-1}$ consisting of those permutations $\pi $ for which
$\pi (1)<\cdots <\pi (k)>\cdots >\pi (m)$.
\ethm
{\bf Proof.}  Multiplying out the terms in the product defining
$C^{(\nu ),m}$, we find that
\begin{eqnarray*}
C^{(\nu ),m}&=&\sum_{s=0}^{m-1}(-1)^{s}\sum_{1\leq i_{1}<\cdots <i_{s}
\leq m-1}\hat{R}_{\nu}(t_{i_{1},m})\cdots \hat{R}_{\nu}(t_{i_{s},m})\\
\hbox{what by Property 4 (in 1.8)}
&=&\sum_{s=0}^{m-1}(-1)^{s}\sum_{1\leq i_{1}<\cdots <i_{s}\leq m-1}
\hat{R}_{\nu}(t_{i_{1},m}\cdots t_{i_{s},m}).
\end{eqnarray*}
The element $\sigma =t_{i_{1},m}t_{i_{2},m}\cdots t_{i_{s},m}$ of
$S_{m}\times S_{1}^{n-m}$ maps $i_{1}$ to $m$, $i_{2}$ to
$m-1$, \dots, and $i_{s}$ to $m-s+1$ and maps the rest in $\{ 1,2,\dots ,m\} $
monotonically increasingly to $\{ 1,2,\dots ,m-s\} $. Moreover it is clear
that the number of inversions $|I(\sigma )|=\sum_{j=1}^{s}|I(t_{i_{j},m})|$
(c.f. Property 4 in 1.8).

The Proposition now follows by setting $\pi =\sigma^{-1}$ and $k=m-s$.
\rem{\bf Remark} 2.1.3.  The Propositions 2.1.1 and 2.1.2 are multiparametric
extensions of Propositions 3. and 4. of [Zag].

\subsection{Bo\v zejko-Speicher type formulas} 

In addition to the, multiplicative in spirit, Zagier type formula for
the inverse of $A^{(\nu )}$ ($\nu $ generic), given in 2.1., one also
has another, additive in spirit, Bo\v zejko-Speicher type formula
(c.f. [BSp1], Lemma 2.6.) which, in the case of symmetric group
$S_{n}$, we shall present here, in slightly different notation,
together with several improvements.

We point out that in Zagier type factorizations
(see Prop.1.8.3) one of the key ingredients was the following coset
decomposition of the symmetric group $S_{n}$ with respect
to its Young subgroup $S_{\{ n-1\} }:=S_{n-1}\times S_{1} :$
$$ S_{n}=S_{\{ n-1\} }\beta_{\{ n-1\} },$$
with $\beta_{\{ n-1\} }=\{ t_{1,n},t_{2,n},\dots ,t_{n,n}\} $ consisting
of distinct coset representatives $t_{k,n}$ $(=$ the cyclic permutation
$ \left( \begin{array}{cccc}
k & k+1 & \cdots & n \\
n & k & \cdots & n-1
\end{array} \right) $). Note that
$\beta_{\{ n-1\} }=\{ g\in S_{n}| g^{-1}(1)<\cdots <g^{-1}(n-1)\} $,
so each $t_{k,n}$ is of smallest
length in the coset $S_{\{n-1\}}t_{k,n}$, it generates, for each $k$,
$1\leq k\leq n$.

Similary we have the left coset decomposition
$$S_{n}=\gamma_{\{ 1\} }S_{\{ 1\} },$$
where $S_{\{ 1\} }:=S_{1}\times S_{n-1}$,
$\gamma_{\{ 1\} }=\{ g\in S_{n}| g(2)<\cdots <g(n)\} =\{ t_{1,1},t_{1,2},
\dots ,t_{1,n-1}\} $.

In general for $J=\{ j_{1}<j_{2}<\cdots <j_{l-1}\} \subseteq
\{ 1,2,\dots ,n-1\} $ let
$S_{J}$ be the Young subgroup of $S_{n}$
$$S_{J}:=S_{j_{1}}\times S_{j_{2}-j_{1}}\times \cdots \times S_{n-j_{l-1}},\
\ S_{\phi }=S_{n}.$$
Note that with such an indexing the Young subgroup $S_{J}$ is generated by all
adjacent transpositions $t_{i}=t_{i,i+1}$, $i\in J^{c}$,
$J^{c}=\{ 1,2,\dots ,n-1\} \setminus J$, (e.g.
$S_{\phi }=S_{n}$ is generated by $t_{1},\dots ,t_{n-1}$), and hence $S_{J}$
is the nontrivial product of the symmetric groups corresponding to the
maximal components of consecutive elements in the complement $J^{c}$.

Then the following is the left coset decomposition :
$$S_{n}=\gamma_{J}S_{J}$$
where $\gamma_{J}=\{ g\in S_{n}| g(1)<g(2)<\cdots <g(j_{1}), g(j_{1}+1)<
\cdots <g(j_{2}),\cdots ,g(j_{l-1}+1)<\cdots <g(n)\} $.

The definition of $\gamma_{J}$ can also be put in the following way
\rem{FACT 2.2.1.} $g\in \gamma_{J}\Leftrightarrow g(1)g(2)\cdots g(n)$ is the
shuffle of the sets \newline
$[1..j_{1}], [j_{1}+1..j_{2}],\dots [j_{l-1}+1,n]\Leftrightarrow $ the
descent set $Des(g)=\{ 1\leq i\leq n-1 | g(i)>g(i+1)\} $ of $g$ is contained
in the set $J$ (c.f. [Sta, pp. 69-70]). (Here $[a..b]$ denotes the set
$\{ a,a+1,\dots ,b\} $.)

Moreover, each $g\in S_{n}$ has the unique factorization $g=a_{J}g_{J}$
with $g_{J}\in S_{J}$ and $a_{J}\in \gamma_{J}$ and with $l(g)=l(a_{J})
+l(g_{J})$.

For arbitrary subset $X\subseteq S_{n}$ we define the matrix $\hat{R}_{\nu }
(X)$ by
$$\hat{R}_{\nu }(X):=\sum_{g\in X}\hat{R}_{\nu }(g)$$
\bthm{PROPOSITION 2.2.2.} Let $\nu $ be a generic weight, $|\nu |=n$. For any
subset
$J=\{ j_{1}<j_{2}<\cdots j_{l-1}\} $ of $\{ 1,2,\dots ,n-1\} $
let $A^{(\nu )}_{J}, \Gamma
^{(\nu )}_{J}$ be the following matrices
$$A^{(\nu )}_{J}=\hat{R}_{\nu }(S_{J}) ,
\Gamma^{(\nu )}_{J}=\hat{R}_{\nu }(\gamma_{J})
.$$
Then the matrix $A^{(\nu )} (=A^{(\nu )}_{\phi })$ of the sesquilinear
form $(\ ,\ )_{{\bf q}}$ (see Prop.1.8.1) has the following factorizations
$$\belowdisplayskip0pt A^{(\nu )}=\Gamma^{(\nu )}_{J}A^{(\nu )}_{J}
$$
$$\belowdisplayskip0pt \Gamma^{(\nu )}_{J}=A^{(\nu )}[A^{(\nu )}_{J}]^{-1}$$
\ethm
{\bf Proof.} By quasimultiplicativity of $\hat{R}_{\nu }$ and FACT 2.2.1
we have $\hat{R}_{\nu }(g)=\hat{R}_{\nu }(a_{J})\hat{R}_{\nu }(g_{J})$.
Hence $A^{(\nu )}=\hat{R}_{\nu }(S_{n})=\hat{R}_{\nu
}(\gamma_{J})\hat{R}_{\nu }(S_{J})=\Gamma^{(\nu )}_{J}A^{(\nu )}_{J}$.

The following formula is the Bo\v zejko-Speicher adaptation of an Euler-type
character formula of Solomon. In the case $W=S_{n}$ it reads as follows :
\bthm{LEMMA 2.2.3.} (c.f.[BSp2] Lemma 2.6) Let $w_{n}=n\dots 2 \: 2 \: 1$ be the
longest permutation in $S_{n}$. Then we have
$$\belowdisplayskip0pt \sum_{J\subseteq \{ 1,2,\dots ,n-1\} }(-1)^{n-1-|J|}\Gamma^{(\nu )}_{J}=
\hat{R}_{\nu }(w_{n})$$
\ethm

For the reader's convenience we include here a variant of the proof
(our notation is slightly different). For any subset $M\subseteq \{ 1,2,\dots
,n-1\} $ we denote by $\delta_{M}$  the subset of $S_{n}$ consisting of all
permutations  $g\in S_{n}$ whose descent set $Des(g)$ is equal to M. Then by
FACT 2.2.1  it is clear that $\gamma_{J}=
\bigcup_{M\subseteq J}\delta_{M}$ (disjoint union), implying that
$$\hat{R}_{\nu }(\gamma_{J})=\sum_{M\subseteq J}\hat{R}_{\nu }(\delta_{M})$$
By the inclusion-exclusion principle we obtain
$$\hat{R}_{\nu }(\delta_{M})=\sum_{J\subseteq M}(-1)^{|M-J|}\hat{R}_{\nu }
(\gamma_{J})$$
By letting $M=\{ 1,2,\dots ,n-1\} (\Rightarrow \delta_{M}=\{ w_{n}\} )$ we
obtain the desired identity.

By combining Prop.2.2.2. and Lemma 2.2.3 we obtain the following relation
among the inverses of matrices $A^{(\nu )}_{J}$'s.
\bthm{PROPOSITION 2.2.4. (Long recursion for the inverse of $A^{(\nu )}$):}
We have
$$[A^{(\nu )}]^{-1}=(\sum_{\phi \neq J\subseteq \{ 1,2,\dots ,n-1\} }
(-1)^{|J|+1}[A^{(\nu )}
_{J}]^{-1})(I+(-1)^{n}\hat{R}_{\nu }(w_{n}))^{-1}$$
\ethm
{\bf Proof.}  By substituting $\Gamma_{J}^{(\nu )}=A^{(\nu )}[A^{(\nu )}_{J}]
^{-1}$ (Prop.2.2.2) into Lemma 2.2.3 and by multiplying by $[A^{(\nu )}]^{-1}$
we obtain
\begin{eqnarray*}
[A^{(\nu )}]^{-1}\hat{R}_{\nu }(w_{n})&=&\sum_{J\subseteq \{ 1,2,\dots ,n-1\} }
(-1)^{n-1-|J|}[A^{(\nu )}_{J}]^{-1} \\
&=&(-1)^{n-1}[A^{(\nu )}_{\phi }]^{-1}+\sum_{\phi \neq J\subseteq
\{ 1,2,\dots ,n-1\} }(-1)^{n-1-|J|}[A^{(\nu )}_{J}]^{-1}
\end{eqnarray*}
But $A^{(\nu )}_{\phi }=A^{(\nu )}$, so the proof follows.
\rem{REMARK 2.2.5.}  Let us associate to each subset $\phi \neq J=\{ j_{1}<j_{2}<
\cdots <j_{l-1}\} \subseteq \{ 1,2,\dots ,n-1\} $ a {\em subdivision} $\sigma (J)$
of the set $\{ 1,2,\dots ,n\} $ into intervals by
$$\sigma (J)= J_{1}J_{2}\cdots J_{l},$$
where $J_{k}=[j_{k-1}+1..j_{k}] (j_{0}=1,j_{l}=n)$. (Here $[a..b]$ denotes
the interval $\{ a,a+1,\dots ,b-1,b\} $ and abbreviate $[a..a] (=\{ a\} )$ to
$[a]$).

The Young subgroup $S_{J}$ can be written as direct product of commuting
subgroups
$$S_{J}=S_{[1..j_{1}]}S_{[j_{1}+1..j_{2}]}\cdots S_{[j_{l-1}+1..n]}
=S_{J_{1}}S_{J_{2}}\cdots S_{J_{l}}$$
where for each interval $I=[a..b]$, $1\leq a\leq b\leq n$ we denote by
$S_{I}=S_{[a..b]}$ the subgroup
of $S_{n}$ consisting of permutations which are the identity on the complement
of $[a.. b]$ (i.e. $S_{[a..b]}=S_{1}^{a-1}\times S_{b-a+1}\times
S_{1}^{n-b}$). By denoting $A^{(\nu )}_{I}=A^{(\nu )}_{[a..b]}:=
\hat{R}_
{\nu }(S_{[a..b]})$, we can rewrite the formula for $[A^{(\nu
)}]^{-1}=[A^{(\nu )}_{[1..n]}]^{-1}$ in Prop.2.2.4. as follows:
$$[A^{(\nu )}_{[1..n]}]^{-1}=(\sum_{\sigma =J_{1}\cdots
J_{l},l\geq 2}(-1)^{l}[A^{(\nu )}_{J_{1}}]^{-1}\cdots [A^{(\nu
)}_{J_{l}}]^{-1})(I+(-1)^{n}\hat{R}_{\nu }(w_{n}))^{-1}\eqno(*)$$
where the sum is over all subdivisions of the set $\{ 1,2,\dots ,n\} $.
Similar formula we can write for $[A^{(\nu )}_{[a..b]}]^{-1}$ for any
nondegenerate interval $[a..b]$, $1\leq a<b\leq n$. Of course if $a=b,
[A^{(\nu )}_{[a..b]}]^{-1}$ is the identity matrix.

Now we shall use an ordering denoted by $<$ on the set $\Sigma_{n}$ of
all subdivisions of the set $\{ 1,2,\dots ,n\} $, called {\em reverse refinement
order}, defined by
$\sigma <\sigma ^{'}$ if $\sigma ^{'}$ is finer than $\sigma $ i.e.
$\sigma ^{'}$ is obtained by subdividing
each nontrivial interval in $\sigma $. The minimal and maximal elements in
$\Sigma_{n}$ are denoted by $\hat 0_{n}(=[1..n])$ and ${\hat
1_{n}}=[1][2]\cdots [n]$. We shall call $(\Sigma_n,<)$ the {\em lattice of
subdivisions} of $\{1,2,\dots, n\}$.
For example we have\newline
$\Sigma_{1}=\{ [1]\}, \Sigma_{2}=\{ [12], [1][2]\} ,
\Sigma_{3}=\{ [123], [1][23], [12][3], [1][2][3]\} ,\newline
\Sigma_{4}=\{ [1234], [123][4],
[12][34], [1][234], [12][3][4], [1][23][4], [1][2][34], [1][2][3][4]\} $.
(Here $[1234]$ denotes the interval $[1..4]=\{ 1,2,3,4\} $ etc.)

\begin{figure}
\caption
{$\Sigma_{4}=$ The lattice of subdivisions of $\{ 1,2,3,4\} $.}
\end{figure}

Now for each interval $I=[a..b], 1\leq a<b\leq n$ we denote by
$w_{I}=w_{[a..b]}:=1\ 2\cdots a-1\ b\ b-1\cdots a\ b+1\cdots n$
the longest permutation in $S_{[a..b]} (=S_{1}^{a-1}\times
S_{b-a+1} \times S_{1}^{n-b})$ and by
$\Psi_{I}^{\nu}=\Psi^{\nu}_{[a..b]}, a<b$ the following matrix
\begin{eqnarray*}
\Psi^{\nu}_{I}&=&\Psi^{\nu}_{[a..b]}:=[I+(-1)^{b-a+1}\hat{R}_{\nu }
(w_{[a..b]})]^{-1}=\frac{1}{\bbox^{\nu}_{[a..b]}}[I-(-1)^{b-a+1}\hat{R}_{\nu
}(w_{[a..b]})]\\
&=&\frac{1}{\bbox_{I}^{\nu}}\Phi_{I}^{\nu}, \qquad \ \Phi_{I}^{\nu}:=
I-(-1)^{|I|}\hat{R}_{\nu}(w_{I})
\end{eqnarray*}
where $\bbox^{\nu}_{[a..b]}$ is the diagonal matrix (agreeing with the
definition of $\bbox^{\nu}_{T}$ given in 1.8):
$$\bbox^{\nu}_{[a..b]}=\bbox^{\nu}_{\{ a,a+1,\cdots ,b\} }
=I-Q^{\nu}_{\{ a,a+1,\dots ,b\} }=I-\prod_{a\leq k<l\leq b}
|Q^{\nu}_{k,l}|^{2}, \ [Q^{\nu}_{k,l}]_{i_{1}\cdots i_{n},i_{1}\cdots i_{n}}=
q_{i_{k}i_{l}}$$
Accordingly, for each subdivision $\sigma =I_{1}I_{2}\cdots I_{l} \in
\Sigma_{n}$ we define
$$\Psi^{\nu}_{\sigma }:=\prod_{j: |I_{j}|\geq 2}\Psi^{\nu}_{I_{j}}$$
and similary for any chain ${\cal C}: \sigma^{(1)}<\cdots <\sigma^{(m)}$ in
$\Sigma_{n}$ we define
$$\Psi^{\nu}_{\cal C}=\Psi^{\nu}_{\sigma^{(m)}}\cdots \Psi^{\nu}_
{\sigma^{(1)}}$$
In the same way we introduce notations $\bbox^{\nu}_{\cal C}$ and $\Phi^{\nu}_
{\cal C}$ and observe that then
$$\Psi^{\nu}_{\cal C}=\frac{1}{\bbox^{\nu}_{\cal C}}\Phi^{\nu}_{\cal C}$$
For example if ${\cal C}: {\hat 0}_{5}=[12345]<[12][345]<[1][2][34][5]<{\hat
1}_{5}$, then
\begin{eqnarray*}
\Psi^{\nu}_{\cal C}&=&\Psi^{\nu}_{\{ 3,4\} }(\Psi^{\nu}_{\{ 1,2\} }\Psi^{\nu}
_{\{ 3,4,5\} })\Psi^{\nu}_{\{ 1,2,3,4,5\} }\\
&=&\frac{1}{\bbox^{\nu}_{\{ 3,4\} }\bbox^{\nu}_{\{ 1,2\} }
\bbox^{\nu}_{\{ 3,4,5\} }\bbox^{\nu}_{\{ 1,2,3,4,5\} }}\Phi^{\nu}_{\{ 3,4\} }
\Phi^{\nu}_{\{ 1,2\} }\Phi^{\nu}_{\{ 3,4,5\} }\Phi^{\nu}_{\{ 1,2,3,4,5\} },
\end{eqnarray*}
for any generic weight $\nu , |\nu |=5$.

Now we can state our first explicit formula for the inverse of $A^{(\nu )}$
in terms of the involutions $w_{I}=w_{[a..b]}, 1\leq a<b\leq n$.
\bthm{THEOREM 2.2.6.} Let $\nu $ be a generic weight, $|\nu |=n$. Then
$$[A^{(\nu )}]^{-1}=\sum_{\cal C }(-1)^{b_{+}({\cal C})+n-1}\Psi^{\nu}_{\cal C}
=\sum_{\cal C}\frac{(-1)^{b_{+}({\cal C})+n-1}}{\bbox^{\nu}_{\cal
C}}\Phi^{\nu}_{\cal C}$$
where the summation is over all chains ${\cal C}: {\hat 0_{n}}=
\sigma^{(0)}<\sigma^{(1)}\cdots <\sigma ^{(m)}<{\hat 1_{n}}$ in the
subdivision lattice $\Sigma_{n}$ and where
$b_{+}({\cal C})$ denotes the total number of nondegenerate intervals
appearing in members of $\cal C $.
\ethm
{\bf Proof.}  The formula follows by iterating the formula (*) in Remark
2.2.5.
\rem{REMARK 2.2.7.}  If we represent chains ${\cal C} : {\hat
0}_{n}=\sigma^{(0)}<\sigma^{(1)} <\cdots <\sigma^{(m-1)}<{\hat 1}_{n}$ of
length $m\geq 1$ as {\em generalized bracketing} (of {\em depth} $m$) of the word $12\cdots
n$ with one pair of brackets for each nondegenerate interval appearing in the
members of $\cal C $ (e.g. ${\hat 0}_{5}
=[12345]<[12][345]<[1][2][34][5]<{\hat 1}_{5}$ is represented as
$[[12][[34]5]]$), then we can write the formula in Thm.2.2.6 as
$$[A^{(\nu )}]^{-1}=\sum_{\beta }(-1)^{b(\beta )+n-1}\Psi^{\nu}_{\beta }
=\sum_{\beta}\frac{(-1)^{b(\beta)+n-1}}{\bbox^{\nu}_{\beta}}\Phi^{\nu}_{\beta}$$
where the sum is over all generalized bracketings of the word $12\cdots n$
and where $b(\beta )$ denotes the number of pairs of brackets in $\beta $
and where $\Psi^{\nu}_{\beta }:=\Psi^{\nu}_{\cal C }$, $\Phi^{\nu}_{\beta}
:=\Phi^{\nu}_{\cal C}$, $\bbox^{\nu}_{\beta}:=\bbox^{\nu}_{\cal C}$ if $\beta $
is associated to the (unique!) chain $\cal C$ in $\Sigma_{n}$ (e.g.
$\Psi^{\nu}_{[[12][[34]5]]}= \Psi^{\nu}_{[3..4]}
(\Psi^{\nu}_{[1..2]}\Psi^{\nu}_{[3..5]})\Psi^{\nu}_{[1..5]}=\Psi^{\nu}_{[1..2]}
\Psi^{\nu}_{[3..4]}\Psi^{\nu}_{[3..5]}
\Psi^{\nu}_{[1..5]}$\newline
$\dsty=\frac{1}{\bbox^{\nu}_{\{ 1,2\} }\bbox^{\nu}_{\{ 3,4\} }
\bbox^{\nu}_{\{ 3,4,5\} }
\bbox^{\nu}_{\{ 1,2,3,4,5\} }}(I-\hat{R}_{\nu }(w_{[1..2]}))(I-\hat{R}_{\nu }
(w_{[3..4]}))(I+\hat{R}_{\nu }(w_{[3..5]}))(I+\hat{R}_{\nu }(w_{[1..5]}))$ .

In particular for Example 1.6.3 ($I=\{ 1,2,3\} , \nu_{1}=\nu_{2}=\nu_{3}=1$) we
have
\begin{eqnarray*}
[A^{123}]^{-1}&=&-\Psi_{[123]}+\Psi_{[[12]3]}+\Psi_{[1[23]]}=\\
&=&\frac{-1}{\bbox_{\{ 1,2,3\} }}(I-\hat{R}_{123}(321))+
\frac{1}{\bbox_{\{ 1,2\} }\bbox_{\{ 1,2,3\} }}
(I+\hat{R}_{123}(213))(I-\hat{R}_{123}(321))+\\
&+&\frac{1}{\bbox_{\{ 2,3\} }
\bbox_{\{ 1,2,3\} }}(I+\hat{R}_{123}(132))(I-\hat{R}_{123}(321)).
\end{eqnarray*}
Similary for $I=\{ 1,2,3,4\}, \nu_{1}=\nu_{2}=\nu_{3}=\nu_{4}=1$ we have
\begin{eqnarray*}
[A^{1234}]^{-1}&=&\Psi_{[1234]}-\Psi_{[1[234]]}-\Psi_{[12[34]]}-
\Psi_{[1[23]4]}-\Psi_{[[12]34]}-\Psi_{[[123]4]}+\\
&+&\Psi_{[[12][34]]}+\Psi_{[[[12]3]4]}+\Psi_{[[1[23]]4]}+\Psi_{[1[[23]4]]}+\Psi_{[1[2[34]]]}
\end{eqnarray*}
(Here we suppresed the upper indices in $\Psi^{123}_{\beta}$
and $\Psi^{1234}_{\beta}$).
\bthm{COROLLARY 2.2.8. (EXTENDED ZAGIER'S CONJECTURE):} For $\nu $ generic,
$|\nu |=n$, for the inverse of the matrix $A^{(\nu)}=A^{(\nu)}({\bf q})$ we
have
$$[A^{(\nu)}]^{-1}\in \frac{1}{\bbox^{\nu}}Mat_{n!}(Z[q_{ij}])\leqno i)$$
where $\bbox^{\nu}$ is the following diagonal matrix
$$\bbox^{\nu}:=\prod_{1\leq a<b\leq n}\bbox^{\nu}_{[a..b]}
=\prod_{1\leq a<b\leq n}(I-Q^{\nu}_{[a..b]})=\prod_{1\leq a<b\leq n}
(I-\prod_{a\leq k\neq l\leq b}Q^{\nu}_{k,l})$$
$$[A^{(\nu)}]^{-1}\in \frac{1}{d_{\nu}}Mat_{n!}(Z[q_{ij}])\leqno i')$$
where $d_{\nu}$ is the following quantity
$$d_{\nu}:=\prod_{\mu\subseteq \nu ,|\mu|\geq 2}\Box_{\mu}
=\prod_{\mu\subseteq \nu ,|\mu|\geq 2}(1-{q}_{\mu})
=\prod_{\mu\subseteq \nu ,|\mu|\geq 2}(1-\prod_{i\neq j\in \mu}
q_{ij})$$
($\Box_{\mu}$ and ${q}_{\mu}$ are the same as in Lemma 1.9.1).

In particular when all $q_{ij}=q$ (Zagier's case) we have from i):
$$[A^{\nu}(q)]^{-1}\in \frac{1}{\delta_{n}(q)}Mat_{n!}(Z[q]) \leqno ii)$$
where
$$\delta_{n}(q)=\prod_{1\leq a<b\leq n}(1-q^{(b-a+1)(b-a)})
=\prod_{k=2}^{n}(1-q^{k(k-1)})^{n-k+1}$$
\ethm
{\bf Proof.} i) follows from Thm 2.2.6 by taking the common denominator
which turns out to be $\bbox^{\nu}=\prod_{1\leq a<b\leq n}\bbox^{\nu}_
{[a..b]}$ because any $\bbox^{\nu}_{[a..b]}$ appears at most once in each
of the denominators $\bbox^{\nu}_{\cal C}$ (and actually appears in at least
one of them).\newline
i') The entries of $\bbox^{\nu}$ are zero or $\bbox^{\nu}_{{\bf i,i}}$
where ${\bf i}=i_{1}\cdots i_{n}$ is any permutation of $\nu $
$(|{\bf i}|=\nu )$ considered as a subset of $I$ (because $\nu $ is
generic!). Since
\begin{eqnarray*}
\bbox^{\nu }_{{\bf i,i}}&=&\prod_{1\leq a<b\leq n}(1-\prod_{a\leq k\neq l\leq b}
q_{i_{k}i_{l}})\\
&=&\prod_{1\leq a<b\leq n}(1-{q}_{\{ i_{a},i_{a+1},\dots ,
i_{b}\} })=\prod_{1\leq a<b\leq n}\Box_{\{ i_{a},i_{a+1},\dots ,i_{b}\} }
\end{eqnarray*}
we see that $\bbox^{\nu}_{{\bf i,i}}$ divides $d_{\nu}$.\newline
ii) Note that in case all $q_{ij}=q$:
\begin{eqnarray*}
\bbox^{\nu}_{{\bf i,i}}&=&\prod_{1\leq a<b\leq n}(1-\prod_{a\leq k\neq l\leq b}
q)=\\
&=&\prod_{1\leq a<b\leq n}(1-q^{(b-a+1)(b-a)})=\prod_{k=2}^{n}(1-q^{k(k-1)})
^{n-k+1}=\delta_{n}(q).
\end{eqnarray*}
This completes the proof of the Extended Zagier's conjecture.
\rem{REMARK 2.2.9.} In [Zag] p.201 Zagier conjectured that $A_{n}(q)^{-1}
\in \frac{1}{\triangle_{n}}Mat_{n!}(Z[q])$, where $\triangle_{n}:=
\prod_{k=2}^{n}(1-q^{k(k-1)})$ and checked this conjecture for $n\leq 5$.
But we found that this conjecture failed for $n=8$
(see Examples to Prop.2.2.18). It seems that our
statement in Corollary 2.2.8 ii) is the right form of a conjecture valid for
all $n$ when all $q_{ij}$ are equal.
\bthm{PROPOSITION 2.2.10.} Let $c_{n}$ be the number ${\hat 0}_{n}-{\hat
1}_{n}$ chains in the subdivision lattice
$\Sigma_{n}$ (i.e. the number of $\Psi $-terms in the formula for $[A^{(\nu )}]
^{-1}$ $\nu $ generic, $|\nu |=n$ in Thm.2.2.6 ), $c_{0}:=0, c_{1}:=1$. Then
$$C(t)=\sum_{n\geq 0}c_{n}t^{n}=\frac{1}{4}(1+t-\sqrt
{1-6t+t^{2}})=t+t^{2}+3t^{3}+11t^{4}+45t^{5}+197t^{6}+\cdots $$
\ethm
{\bf Proof.}  By Remark 2.2.7 this counting is equivalent to the Generalized
bracketing problem of Schr\"oder (1870) (see [Com], p.56).

By expanding the root $(1+u)^{1/2}, u=-6t+t^{2}$ we obtain
$$c_{n}=\sum_{0\leq \nu \leq n/2}(-1)^{\nu }\frac{(2n-2\nu -3)!!}
{\nu !(n-2\nu )!}3^{n-2\nu }2^{-\nu -2}$$

Another formula follows by applying the Lagrange inversion to
$\tilde{C}=\frac{1+2t\tilde{C}^{2}}{1+t}$, where $C=t\tilde{C}$ :
$$c_{n}=\sum_{\nu =0}^{n-1}(-1)^{n-1-\nu }\frac{2^{\nu }}{2\nu +1}
\left( \begin{array}{c}
2\nu +1 \\
\nu
\end{array} \right)
\left( \begin{array}{c}
n+\nu -1 \\
n-\nu -1
\end{array} \right)$$

In fact, the numbers $c_{n}$ can be computed faster via linear
reccurence relation (following from the fact that $C(t)$ is algebraic ):
$$(n+1)c_{n+1}=3(2n-1)c_{n}-(n-2)c_{n-1}, \ \  n\geq 2, \ \ c_{1}=c_{2}=1.$$

Finally, we note that the numbers $q_{n}=2c_{n}$, $n\geq 2$, $q_{1}=1$,
$q_{2}=2$ have yet another interpretation as the numbers of underdiagonal
(except at the ends) paths from $(0,0)$ to $(n,n)$ with step set $\{ (1,0),
(0,1), (1,1)\} $ (c.f. [Com], p.81).

>From this interpretation we get
$$c_{n}=\sum_{r=0}^{n-1}\frac{1}{2n-1-r}
\left( \begin{array}{c}
2n-1-r \\
r, n-r, n-r-1
\end{array} \right ) $$
(cf. [Mo], p.20).

Now we turn our attention to the computation of entries in the inverse of
$A^{(\nu )}$, $\nu $ generic. First we note that any $n!\times n!$ matrix
$A$ can be written as
$$A=\sum_{g\in S_{n}}A(g)R_{n}(g)$$
where $A(g)$ are diagonal matrices defined by $A(g)_{{\bf i,i}}=A_{{\bf
i},g^{-1}\cdot {\bf i}}$ (all ${\bf i}$) ($R_{n}(g)$ is the right regular
representation matrix $R_{n}(g)_{{\bf i,j}}=\delta_{{\bf i},g\cdot {\bf j}}$,
c.f. 1.8).

We call $A(g)$ the {\em g-th diagonal} of $A$.

Hence, if we write
\begin{eqnarray*}
A^{(\nu )}&=&\sum_{g\in S_{n}}A^{(\nu )}(g)R_{\nu }(g),\\{}
[A^{(\nu )}]^{-1}&=&\sum_{g\in S_{n}}[A^{(\nu )}]^{-1}(g)R_{\nu }(g)
\end{eqnarray*}
then by Prop.1.8.1 (in case $\nu $ generic ) we have
$$A^{(\nu )}(g)=Q^{\nu}(g)=\prod_{(a,b)\in I(g^{-1})}Q^{\nu}_{a,b}\  ; \
(Q^{\nu}_{a,b})_{{\bf i,i}}=q_{i_{a}i_{b}}$$
In order to compute $[A^{(\nu )}]^{-1}(g)$ we first write
$$[A^{(\nu )}]^{-1}(g)=\Lambda^{\nu }(g)A^{(\nu )}(g)$$
where $\Lambda^{\nu }(g)$ are yet unknown diagonal matrices.

Similary, for each subset $\emptyset \neq J=\{ j_{1}<j_{2}<
\cdots <j_{l-1}\} \subseteq \{ 1,2,\dots ,n-1\} $ we write
$$[A^{(\nu )}_{J}]^{-1}(g)=\Lambda^{\nu}_{J}(g)A^{(\nu)}_{J}(g)$$
and for any segment $I=[a..b]\subseteq \{ 1,2,\dots ,n\} $
$$[A^{(\nu)}_{I}]^{-1}(g)=\Lambda^{\nu}_{I}(g)A^{(\nu)}_{I}(g)$$
where $\Lambda^{\nu}_{J}(g)$ and $\Lambda^{\nu}_{I}(g)$ are unknown
diagonal matrices.

If $\sigma (J)=J_{1}J_{2}\cdots J_{l}$ is the subdivision
of $\{ 1,2,\dots ,n\} $ (cf. Remark 2.2.5) associated to $J$,
and if $g=g_{1}g_{2}\cdots g_{l} \in S_{J}=S_{J_{1}}S_{J_{2}}\cdots
S_{J_{l}}$, then $\Lambda^{\nu}_{J}(g)=\Lambda^{\nu}_{J_{1}}(g_{1})
\cdots \Lambda^{\nu}_{J_{l}}(g_{l})$.

Let us denote by $S^{>}_{n}$ (resp. $S^{<}_{n}$) the subset
of $S_{n}$ of all elements $g$ such that $g(1)>g(n)$
(resp. $g(1)<g(n)$). It is evident that
$S^{<}_{n}=S^{>}_{n}w_{n}$, $S^{>}_{n}=S^{<}_{n}w_{n}$, where $w_{n}=n n-1
\cdots 2 1$.
\bthm{PROPOSITION 2.2.11.}  The diagonal matrices $\Lambda^{\nu }(g)$ are
real and satisfy the following recurrences :\newline
i)  $\dsty\Lambda^{\nu }(g)=(-1)^{n-1}|Q^\nu(gw_{n})|^{2}\Lambda^{\nu }(gw_{n})$, if
$g \in S^{>}_{n}$\newline
ii) $\dsty\Lambda^{\nu}(g)=\Lambda^{\nu}_{[1..n]}(g)=
\frac{1}{\bbox^{\nu}_{[1..n]}}\sum_{\emptyset \neq J\subseteq
\{ 1,2,\dots ,n-1\} , g\in S_{J}}(-1)^{|J|+1}\Lambda^{\nu}_{J}(g)$,
if $g\in S^{<}_{n}$\newline
ii') $\dsty\Lambda^{\nu }(g)=\frac{1}{\bbox^{\nu}_{[1..n]}}
\sum_{g=g^{'}g^{''}\in S_{k}\times S_{n-k}, 1\leq k\leq
n-1}(Q^{\nu}_{[1..k]})^{[g(1)<g(k)]}
\Lambda^{\nu }_{[1..k]}(g^{'})\Lambda^{\nu }_{[k+1..n]}(g^{''})$, \newline
if $g\in S^{<}_{n}$.

In particular, $[A^{(\nu )}]^{-1}(g)=[A^{(\nu )}]^{-1}(gw_{n})=0$
if both $g$ and $gw_{n}$ are not splittable, i.e. if the minimal Young
subgroup containing g (resp. $gw_{n}$) is equal to $S_{n}$.
\ethm
{\bf Proof.}  By substituting the formula $(I+(-1)^{n}\hat{R}_{\nu }(w_{n}))
^{-1}=\frac{1}{\bbox^{\nu}_{[1..n]}}(I-(-1)^{n}\hat{R}_{\nu }(w_{n}))$
into formula for
$[A^{(\nu )}]^{-1}$ in Prop.2.2.4 we see immediately that for $g\in S^{<}_{n}$
$$[A^{(\nu )}]^{-1}(g)=\sum_{\phi \neq J\subseteq \{1, 2, \dots, n-1\}}(-1)^{|J|+1}
[A^{(\nu )}_{J}]^{-1}(g)\frac{1}{\bbox^{\nu}_{[1..n]}}\eqno (*)$$
(Here we use the fact that $A^{(\nu )}_{J}=\sum_{g\in S_{J}}\hat{R}_{\nu }(g)$
has the inverse of the form $[A^{(\nu )}_{J}]^{-1}=\sum_{g\in S_{J}}\Lambda
^{\nu }_{J}(g)\hat{R}_{\nu }(g)$ and that $g\in S_{J}, J\neq \phi \Rightarrow
g(1)<g(n)$ and $gw_{n}(1)>gw_{n}(n)$.)

Then for $g\in S^{>}_{n}$, again by Prop.2.2.4, we have
$$[A^{(\nu )}]^{-1}(g)\hat{R}_{\nu }(g)=[A^{(\nu )}]^{-1}
(gw_{n})\hat{R}_{\nu }(gw_{n})(-1)^{n-1}\hat{R}_{\nu }(w_{n})$$
implying that
\begin{eqnarray*}
\Lambda^{\nu }(g)\hat{R}_{\nu }(g)&=&(-1)^{n-1}\Lambda^{\nu }(gw_{n})
\hat{R}_{\nu }(gw_{n})\hat{R}_{\nu }(w_{n})\\
&=&(-1)^{n-1}\Lambda^{\nu }(gw_{n})|Q^\nu(gw_{n})|^{2}\hat{R}_{\nu }(g)
\end{eqnarray*}
by Property 3.ii) from 1.8. Thus i) is proved.

The property ii) is immediate from (*) because
$[A^{\nu}_{J}]^{-1}(g)\neq 0 \Rightarrow g\in S_{J}$.
To prove ii') we shall use the following
\bthm{LEMMA 2.2.12. (Short recursion for the inverse of $A^{(\nu )}$):} We
have
$$[A^{(\nu )}]^{-1}=(\sum_{k=1}^{n-1}(-1)^{k-1}[A^{(\nu )}_{\{ k\} }]
^{-1}\hat{R}_{\nu }(w_{[1..k]})(I+(-1)^{n}\hat{R}_{\nu }(w_{n})
)^{-1}$$
where $A^{(\nu )}_{\{ k\} }=\hat{R}_{\nu }(S_{k}\times S_{n-k})
$ is just $A^{(\nu )}_{J}$ when $J=\{ k\} $.
\ethm
{\bf Proof of Lemma 2.2.12.}  For fixed $k$, $1\leq k\leq n-1$ we write every
subset $J=\{ j_{1}<\cdots <j_{l-2}<j_{l-1}=k\} $ as $J^{'}\bigcup \{ k\} $
where $J^{'}=\{ j_{1}<\cdots j_{l-2}\}\subseteq \{ 1,2,\dots ,k-1\} $. Then
$$A^{(\nu )}_{J}=\hat{R}_{\nu }(S_{J^{'}}\times {\bf 1}_{n-k})\cdot \hat{R}
_{\nu }({\bf 1}_{k}\times S_{n-k})$$
Now we compute\newline
$\sum_{max J=k}(-1)^{|J|+1}[A^{(\nu )}_{J}]^{-1}=\sum_{J^{'}\subseteq \{
1,2,\dots ,k-1\} }
(-1)^{|J^{'}|}\hat{R}_{\nu }(S_{J^{'}}\times {\bf 1}_{n-k})^{-1}\hat{R}_{\nu }
({\bf 1}_{k}\times S_{n-k})^{-1} $\newline
$=((-1)^{k-1}\hat{R}_{\nu }(S_{k}\times {\bf 1}_{n-k})^{-1}\cdot \hat{R}_{\nu }
(w_{[1..k]}))\hat{R}_{\nu }({\bf 1}_{k}\times S_{n-k})^{-1}$
(by the Proof of Prop.2.2.4)\newline
$=(-1)^{k-1}[\hat{R}_{\nu }(S_{k}\times {\bf 1}_{n-k})\hat{R}_{\nu }({\bf 1}_{k}\times S_{n-k})]^{-1}\hat{R}_{\nu }(w_{[1..k]})$ \\
$=(-1)^{k-1}[A^{(\nu )}_{\{ k\} }]^{-1}\hat{R}_{\nu }(w_{[1..k]})$ \newline
By summing over $k$, $1\leq k \leq n-1$ and substituting into Prop.2.2.4
we are done.

Now we prove ii'). Let $g\in S^{<}_{n}$. Then by substituting
the formula $(I+(-1)^{n}\hat{R}_{\nu }(w_{n}))^{-1}=\frac{1}{\bbox^{\nu}
_{[1..n]}}(I-(-1)^{n}\hat{R}_{\nu }(w_{n}))$ into Lemma 2.2.12 and comparing
the terms involving $\hat{R}_{\nu }(g)$ in both sides we get
$$\begin{array}{rcl}&&\dsty[A^{(\nu )}]^{-1}(g)R_{\nu }(g)=\Lambda^{\nu }(g)\hat{R}_{\nu}(g)\\
&&\dsty=\sum_{1\leq k\leq n-1}(-1)^{k-1}[A^{(\nu )}_{\{ k\} }]^{-1}
(gw_{[1..k]})R_{\nu }(gw_{[1..k]})\hat{R}_{\nu }
(w_{[1..k]})\\
&&\dsty=\sum_{1\leq k \leq n-1, g=g^{'}g^{''}\in S_{k}\times S_{n-k}}
(-1)^{k-1}\Lambda^{\nu }_{\{ k\} }(g^{'}w_{[1..k]}g^{''})\hat{R}_{\nu }
(g^{'}w_{[1..k]}g^{''})\hat{R}_{\nu }(w_{[1..k]})\\
&&\dsty=\sum_{1\leq k\leq n-1, g=g^{'}g^{''}\in S_{k}\times S_{n-k}}
(-1)^{k-1}\Lambda^{\nu }_{[1..k]}(g^{'}w_{[1..k]})\Lambda^{\nu }_{[k+1..n]}
(g^{''})|Q(g^{'}w_{[1..k]})|^{2}\hat{R}_{\nu }(g^{'} g^{''})
\end{array}\eqno(**)$$
(by Prop.1.8.2).

Now, if $g^{'}(1)<g^{'}(k)$ then $g^{'}w_{[1..k]}(1)>g^{'}w_{[1..k]}(k)$ so
by i) we  have
$$\Lambda^{\nu }_{[1..k]}(g^{'}w_{[1..k]})=(-1)^{k-1}|Q^{\nu}(g^{'})|^{2}
\Lambda^{\nu }_{[1..k]}(g^{'}).$$
Then $(-1)^{k-1}\Lambda^{\nu }_{[1..k]}(g^{'}w_{[1..k]})
|Q^{\nu}(g^{'}w_{[1..k]})|^{2}=|Q^{\nu}(w_{[1..k]})|^{2}
\Lambda^{\nu }(g^{'})$. Similary, if $g^{'}(1)>g^{'}(k)$ then $g^{'}w_{[1..k]}
(1)>
g^{'}w_{[1..k]}(k)$, so by i) we have $(-1)^{k-1}\Lambda^{\nu }_{[1..k]}
(g^{'}w_{[1..k]})
|Q^{\nu}(g^{'}w_{[1..k]})|^{2}=\Lambda^{\nu }(g^{'})$.
By substituting these two formulas
into $(**)$ we get
$$\Lambda_{[1..n]}^{\nu }(g)=\frac{1}{\bbox^{\nu}_{[1..n]}}\sum_{1\leq k
\leq n-1, g=g^{'}g^{''}\in S_{k}\times
S_{n-k}}|Q^{\nu}(w_{[1..k]})|^{2[g^{'}(1)<g^{'}(k)]}\Lambda^{\nu }_{[1..k]}
(g^{'})\Lambda^{\nu }_{[k+1..n]}(g^{''}).$$
Finally we use that $|Q^{\nu}(w_{[1..k]})|^{2}=\prod_{1\leq a<b\leq k}
Q^{\nu}_{\{ a,b\} }=Q^{\nu}_{\{ 1,2,\dots ,k\} }=Q^{\nu}_{[1..k]}$.
This completes the proof of Proposition 2.2.11.
\bthm{COROLLARY 2.2.13.}  With notations of Remark 2.2.7 and Proposition
2.2.11 we have the following formulas for the diagonal entries of the inverse
of $A^{\nu}$, $\nu $ generic, $|\nu |=n$.
$$[A^{(\nu )}]^{-1}(id)=\sum_{\beta }\frac{(-1)^{b(\beta)+n-1}}
{\bbox^{\nu}_{\beta}}\leqno i)$$
where the sum is over all generalized bracketings $\beta $ of the word
$12\cdots n$, which have outer brackets.
$$[A^{(\nu )}]^{-1}(id)=\frac{1}{\Box^{\nu}_{[1..n]}}\sum_{\beta }
\frac{Q^{\nu}_{\beta}}{\Box^{\nu}_{\beta}}\leqno i')$$
where the sum is over all generalized bracketings $\beta $ of the word
$12\cdots n$, which don't have outer brackets and where $Q^{\nu}_{\beta}$
is defined, analogously as $\bbox^{\nu}_{\beta}$,
to be the product of $Q^{\nu}_{[a..b]}$ over all bracket pairs in $\beta $.
\ethm
{\bf Proof.} i) follows from Remark 2.2.7 because $\hat{R}_{\nu }$-terms
contribute only to nondiagonal entries .\newline
i') follows by iterating Proposition 2.2.11 $ii)^{'}$ in the case $g=id$ and
using that $[A^{(\nu )}]^{-1}(id)=\Lambda^{\nu }(id)A^{(\nu )}(id)=
\Lambda^{\nu }(id)Q^{\nu }(id)=\Lambda^{\nu }(id)$.

In particular if $I=\{ 1,2\} ,\nu_{1}=\nu_{2}=1$, we have
$\Lambda^{12}(id)=[A^{12}]^{-1}(id)=\frac{1}{\bbox_{\{ 1,2\} }}$.

In Example 1.6.3 ($I=\{ 1,2,3\} ,\nu_{1}=\nu_{2}=\nu_{3}=1$)
we have
\begin{eqnarray*}
\Lambda^{123}(id)=[A^{123}]^{-1}(id)&=&\frac{-1}{\bbox_{\{ 1,2,3\} }}+\frac{1}
{\bbox_{\{ 1,2\} }\bbox_{\{ 1,2,3\} }}+\frac{1}{\bbox_{\{ 2,3\} }\bbox_{\{ 1,2,3\} }}\\
&=&\frac{1}{\bbox_{\{ 1,2,3\} }}(1+\frac{Q_{\{ 1,2\} }}{\bbox_{\{ 1,2\} }}
+\frac{Q_{\{ 2,3\} }}{\bbox_{\{ 2,3\} }})\\
\end{eqnarray*}
Similarly for $I=\{ 1,2,3,4\} ,\nu_{1}=\nu_{2}=\nu_{3}=\nu_{4}=1$ we have
\begin{eqnarray*}
\Lambda^{1234}(id)&=&[A^{1234}]^{-1}(id)=\\
&=&\frac{1}{\bbox_{1234}}\left\{ 1-\frac{1}{\bbox_{12}}
-\frac{1}{\bbox_{23}}-\frac{1}{\bbox_{34}}+\frac{1}{\bbox_{12}\bbox_{34}}\right.\\
&+&\left.(-1+\frac{1}{\bbox_{12}}+\frac{1}{\bbox_{23}})\frac{1}{\bbox_{123}}+(-1+\frac{1}
{\bbox_{23}}+\frac{1}{\bbox_{34}})\frac{1}{\bbox_{234}}\right\}\\[0pt\pagebreak]
&=&\frac{1}{\bbox_{1234}}\left\{ 1+\frac{Q_{12}}{\bbox_{12}}+\frac{Q_{23}}{\bbox_{23}}
+\frac{Q_{34}}{\bbox_{34}}+\frac{Q_{12}Q_{34}}{\bbox_{12}\bbox_{34}}\right.\\
&+&\left.(1+\frac{Q_{12}}
{\bbox_{12}}+\frac{Q_{23}}{\bbox_{23}})\frac{Q_{123}}{\bbox_{123}}+(1+
\frac{Q_{23}}{\bbox_{23}}+\frac{Q_{34}}{\bbox_{34}})\frac{Q_{234}}{\bbox_{234}}
\right\}
\end{eqnarray*}
(Here we abbreviated $Q_{\{ 1,2\} }, Q_{\{ 2,3,4\} }$ to $Q_{12},Q_{234}$
etc.).

If we take all $q_{ij}=q$ (Zagier's case), then we obtain easily that
$$[A_{3}(q)]^{-1}(id)=\frac{1+q^{2}}{(1-q^{2})(1-q^{6})}I$$
$$[A_{4}(q)]^{-1}(id)=\frac{1+2q^{2}+q^{4}+2q^{6}+q^{8}}{(1-q^{2})
(1-q^{6})(1-q^{12})}I$$
which agree with Zagier's computations.
\rem{REMARK 2.2.14.} The formula i') in Corollary 2.2.13 can be interpreted
also as a regular language expression for closed walks in the weighted
digraph (a Markov chain) ${\cal D}^{\nu }$ on the symmetric group $S_{n}$
where the adjacency matrix $A({\cal D}^{\nu })$ is given by nondiagonal entries of $A^{(\nu )}$ multiplied by -1, i.e. $A({\cal D}^{\nu })=-(A^{(\nu )}-I)$.
Then the walk generating matrix function of ${\cal D}^{\nu }$ is nothing but
the inverse of $A^{(\nu )}$ because $W({\cal D}^{\nu })=(I-A({\cal D}^{\nu }))
^{-1}=[A^{(\nu )}]^{-1}$. For example, we have
\begin{eqnarray*}
W({\cal D}^{123})_{closed}&=&[A^{123}]^{-1}(id)=Q_{\{ 1,2,3\} }^{\hphantom{[1..2]}*}
(I+Q_{\{ 1,2\} }^{\hphantom{[1..2]}+}+Q_{\{ 2,3\} }^{\hphantom{[1..2]}+})\\
W({\cal D}^{1234})_{closed}&=&[A^{1234}]^{-1}(id)=Q^{\hphantom{[1..2]}*}_{[1..4]}\left\{ 1+Q^{\hphantom{[1..2]}+}_{[1..2]}+Q^{\hphantom{[1..2]}+}_{[2..3]}
+Q^{\hphantom{[1..2]}+}_{[3..4]}+Q^{\hphantom{[1..2]}+}_{[1..2]}Q^{\hphantom{[1..2]}+}_{[3..4]}\right.\\
&+&\left.(1+Q^{\hphantom{[1..2]}+}_{[1..2]}+Q^{\hphantom{[1..2]}+}_{[2..3]})Q^{\hphantom{[1..2]}+}_{[1..3]}+(1+Q^{\hphantom{[1..2]}+}_{[2..3]}
+Q^{\hphantom{[1..2]}+}_{[3..4]})Q^{\hphantom{[1..2]}+}_{[2..4]}\right\}
\end{eqnarray*}
in the familiar formal language notation ($x^{*}=\frac{1}{1-x}, x^{+}=
\frac{x}{1-x}$).
\rem{REMARK 2.2.15.}  Besides the formulas for $c_{n}=$ total number of
$\Psi_{\cal C}$-terms in the formula for $[A^{(\nu )}]^{-1}$ (Theorem 2.2.6)$=$ total number
of $\bbox^{\nu}_{\beta }$-terms in the formula for $[A^{(\nu )}]^{-1}(id)$
(Corollary 2.2.13 ) we can also give the formulas for the numbers
$c_{n,k}:=Card\ {\cal C}_{n,k}$ ($n\geq 2, 1\leq k\leq n-1$ or $n=1$ and
$k=0$) where ${\cal C}_{n,k}:=$all generalized bracketings of the word
$12...n$ which have outer brackets (surrounding the entire word
$12...n$), having all together k pairs of brackets, if $n\geq 2$.\newline
(i.e. $c_{n,k}=$ number of terms in $[A^{(\nu )}]^{-1}(id)$ having k
$\bbox $-factors).

For example $c_{3,1}=1, c_{3,2}=2, c_{4,1}=1, c_{4,2}=5, c_{4,3}=5$.
\bthm{LEMMA 2.2.16.}We have\newline
i) $$C(t,z)=t+\sum_{n\geq 2, 1\leq k\leq n-1}c_{n,k}t^{n}z^{k}=
\frac{1+t-\sqrt{(1-t)^{2}-4tz}}{2(1+z)}$$
ii) $$c_{n,k}=\frac{1}{n}\left(\begin{array}{c}
                                                 n+k-1 \\
                                                 k
                                                 \end{array} \right)
                                                 \left(\begin{array}{c}
                                                 n-2 \\
                                                 k-1
                                                 \end{array} \right) , n\geq 2, 1\leq k\leq
n-1, c_{1,0}=1.$$
\ethm
{\bf Proof.} We first observe that each bracketing $\beta \in {\cal
C}_{n,k}$ can be viewed as a word $w=w(\beta )\in \{ x,y,\bar{y}\}^{*}
$ (by replacing each left bracket $[$ by y, each right bracket $]$
by $\bar{y}$ and each of the letters $1,2,...,n$ by $x$) such that $|w|
_{y}=k, |w|_{x}=n$. Then for the language ${\cal C}={\cal C}^{'}
+{\cal C}^{''}$ where ${\cal C}^{'}:={\cal C}_{1,0}=x, {\cal C}^{''}
=\bigcup_{n\geq 2, 1\leq k\leq n-1}{\cal C}_{n,k}$ we obtain the
following language equation
$${\cal C}=x+y{\cal C}{\cal C}^{+}\bar{y}\eqno(*)$$
where for any alphabet $A$ we denote by $A^{+}$ the set of
all nonempty words over A, i.e. $A^{+}=A+A^{2}+A^{3}+\cdots
=\frac{A}{1-A}$. By letting $x=t, y=z, \bar{y}=1$, where t and z
commute we obtain from (*) the following quadratic equation
for the corresponding generating function $C=C(t,z)=t+\sum_{n\geq 2,
k\leq n-1}c_{n,k}t^{n}z^{k}$:
$$(1+z)C^{2}-(1+t)C+t=0$$
from which i) follows immediately.

The formula ii) for the coefficients $c_{n,k}$ follows by
Lagrange inversion applied to the following (equivalent) equation for $C$:
$$C=\frac{t}{1-z\frac{C}{1-C}}$$
(Note that Lagrange inversion applied to $C=\frac{t+(1+z)C^{2}}
{1+t}$ would give a refinement of the second formula for $c_{n}$
(see Prop. 2.2.10) with $2^{\nu }$ replaced by $\nu \choose k+1$, but our
formula in ii) is shorter). This proves the Lemma.

Note that $c_{n,n-1}=\frac{1}{n}{2n-2 \choose n-1}$ is just the n-th Catalan
number (cf. [Com].p.53)
\bthm{COROLLARY 2.2.17.}  The numbers $c_{n,k}$ of terms in $[A^{(\nu )}]^{-1}
(id)$ ($\nu $ generic, $|\nu |=n$) having k $\bbox $-factors
(in Corollary 2.2.13) or the numbers of regular expression monomials
of degree k in $W({\cal D}^{(\nu )})_{closed}$ (in Remark 2.2.14) are
the coefficients of the following Catalan-Schr\"oder polynomials :
$$P_{n}(z)=\sum_{k=1}^{n-1}\frac{1}{n}\left(\begin{array}{c}
n+k-1 \\
k
\end{array} \right)
\left(\begin{array}{c}
n-2 \\
k-1
\end{array} \right) z^{k}, n\geq 2, P_{1}(z)=1$$
i.e. $c_{n,k}=[z^{k}]P_{n}(z)$.

Note that $c_{n}=P_{n}(1)=\sum_{k=1}^{n-1}\frac{1}{n}{n+k-1 \choose k}
{n-2 \choose k-1}$ is yet another formula for number $c_{n}$ of
${\hat0_n-\hat1_n}$ chains in the subdivision lattice $\Sigma_{n}$.
\ethm

Now we turn our attention to computing a general entry of the inverse  of
$A^{\nu }$, $\nu $ generic, $|\nu |=n$.

Let $g\in S_{n}^{<}$(i.e. $ g(1)<g(n)$) be given. Let $J(g)=\{ j_{1}<j_{2}<
\cdots <j_{n(g)-1}\} \subset \{ 1,2,\dots ,n-1\} $ be such that $S_{J(g)}$ is
the minimal Young subgroup of $S_{n}$ containing g.
It is clear that $J(g)$ can be given explicitly as
$$J(g)=\{ 1\leq j\leq n-1 | g(1)+g(2)+\cdots +g(j)=1+2+\cdots +j\} $$
Then by $\sigma (g)=J_{1}J_{2}\cdots J_{n(g)} \in \Sigma_{n}$ we denote
the subdivision associated to the set $J(g)$ i.e
$$J_{1}=J_{1}(g):=[1..j_{1}], J_{2}=J_{2}(g):=[j_{1}+1..j_{2}],\cdots ,
J_{n(g)}:=J_{n(g)}(g)=[j_{n(g)-1}+1..n]$$
and by $g=g_{1}g_{2}\cdots g_{n(g)}$ we denote the corresponding factorization
of $g$ with $g_{k}\in S_{J_{k}(g)}, 1\leq k\leq n(g)$.
By noting that $g\in S_{J}\Leftrightarrow J\subseteq J(g)$, we can
rewrite the formula Prop.2.2.11 ii) as follows
\begin{eqnarray*}
\Lambda^{\nu}(g)&=&\Lambda^{\nu}_{[1..n]}(g)=\frac{1}{\bbox^{\nu}_{[1..n]}}
\sum_{\emptyset \neq J\subseteq J(g)}(-1)^{|J|+1}\Lambda^{\nu}_{J}(g)\\
&=&\frac{1}{\bbox^{\nu}_{[1..n]}}\sum_{\emptyset \neq K\subseteq \{ 1,2,\dots
,n(g)-1\} }(-1)^{|K|+1}\Lambda^{\nu}_{J(K)}(g)
\end{eqnarray*}
where $J(K):=\{ j_{k}| k\in K\} \subseteq \{ 1,2,\dots ,n-1\} $.
(Note that if $J(g)=\emptyset $ ($\Rightarrow g$ and $gw_{n}$ are not
splittable), then $\Lambda^{\nu}(g)=0$ by this formula too.)

In terms of subdivisions this can be viewed as a recursion formula:
$$\Lambda_{[1..n]}^{\nu }(g)=\frac{1}{\bbox^\nu_{[1..n]}}\sum_{\tau =
K_{1}K_{2}\cdots K_{l}\in \Sigma_{n(g)}, l\geq 2}(-1)^{l}\Lambda_{I(K_{1})}
^{\nu }(g_{K_{1}})\cdots \Lambda_{I(K_{l})}^{\nu }(g_{K_{l}})\eqno (*)$$
where $I(K_{s}):=\bigcup_{k\in K_{s}}J_{k}(g)$, $g_{K_{s}}:=\prod_{k\in
K_{s}}g_{k}, s=1,...,l$.

By iterating this recursion formula (*) (as in Theorem 2.2.6, Remark 2.2.7,
Corollary 2.2.13) we obtain
$$\Lambda_{[1..n]}^{\nu }(g)=(\sum_{\beta }(-1)^{b(\beta )+n(g)-1}\tilde{\Psi }
_{\beta })\Lambda_{J_{1}(g)}^{\nu }(g_{1})\cdots \Lambda^{\nu }
_{J_{n(g)}(g)}(g_{n(g)})\eqno(**)$$
where $\beta $ run over all generalized bracketings of the word $12\cdots
n(g)$ which have outer brackets and where each bracket pair $[a..b]$, $1
\leq a<b\leq n(g)$, we set
$$\tilde{\Psi}_{[a..b]}:=\frac{1}{\bbox_{J_{a}\bigcup J_{a+1}\bigcup
\cdots \bigcup J_{b}}}$$
($b(\beta):=$number of bracket pairs in $\beta $). Thus the expression in the
parentheses can be viewed as a {\em "thickened" identity coefficient}
$$\Lambda^{12\cdots n(g)}(id)|^{\nu }_{1\rightarrow J_{1}, 2\rightarrow
J_{2},\cdots ,n(g)\rightarrow J_{n(g)}}$$
which we shall denote by
$$\Lambda^{\nu }_{\sigma (g)}=\Lambda^{\nu}
_{J_{1}(g)J_{2}(g)\cdots J_{n(g)}(g)}:=\Lambda^{12\cdots n(g)}(id)
|_{1\rightarrow J_{1},2\rightarrow J_{2},\dots ,n(g)\rightarrow J_{n(g)}}.$$
(In particular we can now write $\Lambda^{\nu}_{[1..n]}(id)$
also as $\Lambda^{\nu}_{[1][2]\cdots [n]}$).

As an example for this notation we take $g=4 1 3 2 5 7 8 6$. Then
$\sigma (g)=[1..4][5][6..8]$ i.e $J_{1}(g)=[1..4], J_{2}(g)=[5],
J_{3}(g)=[6..8]$. So
$$\Lambda^{\nu}_{[1..4][5][6..8]}=\Lambda^{123}(id)|^{\nu}
_{1\rightarrow [1..4], 2\rightarrow [5], 3\rightarrow [6..8]}
=\frac{1}{\bbox^{\nu}_{[1..8]}}(-1+\frac{1}{\bbox^{\nu}_{[1..5]}}
+\frac{1}{\bbox^{\nu}_{[5..8]}})$$
(c.f. Corollary 2.2.13).

Now we have one more observation concerning the formula ($**$). To each
nonzero factor $\Lambda^{\nu}_{J_{k}(g)}(g_{k}), 1\leq k\leq n(g)$ in
($**$) we can apply Prop. 2.2.11 i) because $g_{k}$, being a minimal
Young factor
of $g$, is not splittable and hence \newline
$g_{k}(j_{k-1}+1)>g_{k}(j_{k})$ (otherwise $g_{k}w_{J_{k}}$ would also
be nonsplittable $\Rightarrow \Lambda^{\nu}_{J_{k}(g)}(g_{k})=0$)
$$\Lambda^{\nu}_{J_{k}(g)}(g_{k})=(-1)^{|J_{k}(g)|-1}|Q^{\nu}(g_{k}
w_{J_{k}(g)})|^{2}\Lambda^{\nu}_{J_{k}(g)}(g_{k}w_{J_{k}(g)})$$
By substituting this into ($**$) we obtain the following algorithm
for computing the diagonal matrices $\Lambda^{\nu}(g)$ describing
the inverse of $A^{(\nu )}$\newline
(recall that $[A^{(\nu )}]^{-1}=
\sum_{g\in S_{n}}\Lambda^{\nu}(g)\hat{R}(g)$).
\bthm{PROPOSITION 2.2.18. (An algorithm for $\Lambda^{\nu}(g)$, $\nu $
generic, $|\nu |=n$).}  For $g\in S_{n}$ we have
$$\Lambda^{\nu}_{[1..n]}(g)=(-1)^{n-n(g)}\Lambda^{\nu}_{\sigma (g)}
|Q^{\nu}(g^{'})|^{2}\Lambda^{\nu}_{J(g)}(g^{'})$$
where $g^{'}:=gw_{J(g)}$ ($w_{J(g)}=$ the maximal element in the
minimal Young subgroup $S_{J(g)}$ containing $g$). A similar statement holds
true if we replace $[1..n]$ by any interval $[a..b], 1\leq a\leq b\leq n$.
\ethm
{\bf Proof.}  If $g(1)<g(n)$ this is what we get from ($**$).

If $g(1)>g(n)$, then $J(g)=\emptyset , S_{J(g)}=S_{n}, w_{J(g)}=
n n-1 \dots 2 1=w_{n}, n(g)=1, \sigma (g)=[1..n], \Lambda^{\nu}
_{\sigma(g)}=\Lambda^{1}(id)|_{1\rightarrow [1..n]}=I, g^{'}=gw_{J(g)}=
gw_{n}$, so what we needed to prove is just the claim in Prop.2.2.11 i).
The Prop.2.2.18 is proved.

To illustrate this algorithm we take $g=41325786$ ($\nu $
can be any generic weight, $|\nu |=8$) for which $J(g)=\{ 4,5\} ,
J_{1}(g)=[1..4], J_{2}(g)=[5], J_{3}(g)=[6..8], n(g)=3, n=8, w_{J(g)}
=43215876, g^{'}=gw_{J(g)}=23145687, Q^{\nu}(g^{'})=Q^{\nu}_{1,2}
Q^{\nu}_{1,3}Q^{\nu}_{7,8},$\newline
$ |Q^{\nu}(g^{'})|^{2}=Q^{\nu}(g^{'}) Q^{\nu}(g^{'})^{*}=Q^{\nu}_{\{ 1,2\}
}Q^{\nu}_{\{ 1,3\} }Q^{\nu}_{\{7,8\} }$. Then the first step of our algorithm
gives
\begin{eqnarray*}
&&\Lambda^{\nu}_{[1..8]}(g)=\Lambda^{\nu}_{[1..8]}(41325786)=\\
&&=(-1)^{8-3}\Lambda^{\nu}_{[1..4][5][6..8]}Q^{\nu}_{\{ 1,2\} }
Q^{\nu}_{\{ 1,3\} }Q^{\nu}_{\{ 7,8\} }\Lambda^{\nu}_{[1..4]}(2314)
\Lambda^{\nu}_{[5]}(5)\Lambda^{\nu}_{[6..8]}(687).
\end{eqnarray*}
In the second step of our algorithm we compute\newline
$\Lambda^{\nu}_{[1..4]}(2314)=(-1)^{4-2}\Lambda^{\nu}_{[1..3][4]}
Q^{\nu}_{\{ 2,3\} }\Lambda^{\nu}_{[1..3]}(132)\Lambda^{\nu}_{[4]}(4)$\newline
$\Lambda^{\nu}_{[6..8]}(687)=(-1)^{3-2}\Lambda^{\nu}_{[6][7..8]}
\Lambda^{\nu}_{[6]}(6)\Lambda^{\nu}_{[7..8]}(78)$\newline
In the third (and the final) step we need only to compute
$$\Lambda^{\nu}_{[1..3]}(132)=(-1)^{3-2}\Lambda^{\nu}_{[1][2..3]}
\Lambda^{\nu}_{[1]}(1)\Lambda^{\nu}_{[2..3]}(23).$$
Since $\Lambda^{\nu}_{[7..8]}(78)=\Lambda^{\nu}_{[7][8]}, \Lambda^{\nu}
_{[2..3]}(23)=\Lambda^{\nu}_{[2][3]}, (Q^{\nu}_{\{ 1,2\} }Q^{\nu}_{\{ 1,3\} })
Q^{\nu}_{\{ 2,3\} }=Q^{\nu}_{[1..3]}, \Lambda^{\nu}_{[1]}(1)=\cdots
=\Lambda^{\nu}_{[8]}(8)=I$, we finally obtain
$$\Lambda^{\nu}_{[1..8]}(41325786)=-\Lambda^{\nu}_{[1..4][5][6..8]}
\Lambda^{\nu}_{[1..3][4]}\Lambda^{\nu}_{[1][2..3]}\Lambda^{\nu}
_{[2][3]}\Lambda^{\nu}_{[6][7..8]}\Lambda^{\nu}_{[7][8]}Q^{\nu}_{[1..3]}
Q^{\nu}_{[7..8]}.$$

As a general example we take $g=w_{J}$ where $J=\{ j_{1}<\cdots <j_{l-1}\} $
is an arbitrary subset of $\{ 1,2,\dots ,n-1\} $. Here $n(g)=l$ and $g^{'}
=id$, so by one application of our algorithm we obtain
$$\Lambda^{\nu}_{[1..n]}(w_{J})=(-1)^{n-l}\Lambda^{\nu}_{J_{1}J_{2}
\cdots J_{l}}\Lambda^{\nu}_{J_{1}}(id)\Lambda^{\nu}_{J_{2}}(id)\cdots
\Lambda^{\nu}_{J_{l}}(id)$$
where $J_{1}=[1..j_{1}], J_{2}=[j_{1}+1..j_{2}], \dots ,J_{l}=[j_{l-1}+1..n]$.

In particular for $n=8$, $J=\{ 4\} $ we obtain
\begin{eqnarray*}
\Lambda^{\nu}_{[1..8]}(43218765) &=&(-1)^{8-2}\Lambda^{\nu}_{[1..4][5..8]}
\Lambda^{\nu}_{[1..4]}(1234)\Lambda^{\nu}_{[5..8]}(5678)\\
&=&\frac{1}{\bbox^{\nu}_{[1..8]}}\Lambda^{\nu}_{[1][2][3][4]}\Lambda^{\nu}
_{[5][6][7][8]}
\end{eqnarray*}
In Zagier's case, when all $q_{ij}=q$, we would then have (c.f. Examples
to Cor. 2.2.13)
$$\Lambda^{\nu}_{[1..8]}(43218765)=\frac{1}{1-q^{7\cdot 8}}\frac{
(1+2q^{2}+q^{4}+2q^{6}+q^{8})^{2}}{(1-q^{1\cdot 2})^{2}(1-q^{2\cdot 3})^{2}
(1-q^{3\cdot 4})^{2}}I$$
But the denominator $D_{8}$ of this expression does not divide Zagier's
$\triangle_{8}=(1-q^{2\cdot 1})(1-q^{3\cdot 2})(1-q^{4\cdot 3})(1-q^{5\cdot 4})
(1-q^{6\cdot 5})(1-q^{7\cdot 6})(1-q^{8\cdot 7})$. Namely $\triangle_{8}
/D_{8}=(1-q^{4\cdot 5})(1-q^{5\cdot 6})(1-q^{6\cdot 7})/(1-q^{1\cdot 2})
(1-q^{2\cdot 3})(1-q^{3\cdot 4})$ is not a polynomial due to the factor
$1-q^{2}+q^{4}$ in the denominator. This computation shows
that the original Zagier's conjecture (c.f. Remark 2.2.9) fails for $n=8$.

Now we return to our agorithm. We shall show now that it is somewhat
better to combine two steps of our algorithm into one step. This can be
observed already in our illustrative example ($g=41325786$) where after the
second step the "unrelated factors" $Q^{\nu}_{\{ 1,2\} }$ and $Q^{\nu}_{\{
1,3\} }$ from the first step were completed, with the factor $Q^{\nu}_{\{
2,3\} }$, into a "nicer" term $Q^{\nu}_{[1..3]}$ having a contiguous indexing set.
Fortunately this holds in general, but first we need more notations to state
the results.

To each permutation $g\in S_{n}$ we can associate a sequence of permutations
$g, g^{'}, g^{''}, \dots $, where $g^{(k+1)}$ is obtained from $g^{(k)}$
by reversing all minimal Young factors in $g^{(k)}$ i.e $g^{'}=gw_{J(g)},
g^{''}=gw_{J(g^{'})},\dots ,g^{(k+1)}=(g^{(k)})^{'}=g^{(k)}w_{J(g^{(k)})},
\dots $.

We shall call this sequence a {\em Young sequence} of $g$.

We call $g$ {\em tree-like} if $g^{(k)}=id$ for some $k$, and by {\em depth
of} $g$ we call the minimal such $k$.

Besides the notation $\Lambda^{\nu}_{\sigma (g)}=\Lambda^{\nu}_{J_{1}(g)
J_{2}(g)\cdots J_{n(g)}(g)}$, where $\sigma (g)=J_{1}(g)\cdots J_{n(g)}(g)$
is the subdivision of $\{ 1,2,\dots ,n\} $ associated to the minimal
Young subgroup $S_{J(g)}$ containing $g$ we need a relative version
$\Lambda^{\nu}_{\sigma (g^{'}):\sigma (g)}$ which we define by
$$\Lambda^{\nu}_{\sigma (g^{'}):\sigma (g)}:=\Lambda^{\nu}
_{\sigma (g^{'}|J_{1}(g))}\Lambda^{\nu}_{\sigma (g^{'}|J_{2}(g))}\cdots
\Lambda^{\nu}_{\sigma (g^{'}|J_{n(g)}(g))}$$
For example when $g=41325786 (\Rightarrow g^{'}=23145687)$, $J_{1}(g)
=[1..4], J_{2}(g)=[5], J_{3}(g)=[6..8]$, we have
$$\Lambda^{\nu}_{\sigma (g^{'}):\sigma (g)}=\Lambda^{\nu}_{[123][4]}
\Lambda^{\nu}_{[5]}\Lambda^{\nu}_{[6][7..8]}$$
Also, besides the notation, for $T\subseteq \{ 1,2,\dots ,n\} ,
Q^{\nu}_{T}=\prod_{a,b\in T,a\neq b}Q^{\nu}_{a,b}$
(introduced in 1.8), we define for any
subdivision $\sigma =J_{1}J_{2}\cdots J_{l}$ of $\{ 1,2,\dots ,n\} $:
$$Q^{\nu}_{\sigma }:=Q^{\nu}_{J_{1}}Q^{\nu}_{J_{2}}\cdots Q^{\nu}_{J_{l}}$$
For example:
$$Q^{\nu}_{[1..3][4][5][6][7..8]}=Q^{\nu}_{[1..3]}Q^{\nu}_{[4]}Q^{\nu}_{[5]}
Q^{\nu}_{[6]}Q^{\nu}_{[7..8]}=Q^{\nu}_{[1..3]}Q^{\nu}_{[7..8]}.$$
\bthm{PROPOSITION 2.2.19. (Fast algorithm for $\Lambda^{\nu}(g)$, $\nu $ generic,
$|\nu |=n$):} With the notations above we have
$$\Lambda^{\nu}_{[1..n]}(g)=(-1)^{n(g)+n(g^{'})}\Lambda^{\nu}_{\sigma (g)}
\Lambda^{\nu}_{\sigma (g^{'}):\sigma (g)}Q^{\nu}_{\sigma (g^{'})}\Lambda^{\nu}
_{J(g^{'})}(g^{''})$$
($n(g)=$ the number of minimal Young factors of $g$)
\ethm
{\bf Proof.}  The proof consists in combining together the first two steps of
the algorithm in Proposition 2.2.18.

First we note that in the unique factorization of $g=g_{1}g_{2}\cdots g
_{n(g)}\in S_{J(g)}=S_{J_{1}(g)}S_{J_{2}(g)}\cdots S_{J_{n(g)}(g)}$
(here $J_{k}(g)'s$ denote intervals associated to the set $J(g)\subset
[1\dots n-1]$(cf. Remark 2.2.5)) w.r.t. its minimal
Young subgroup. This implies that
$$g^{'}=gw_{J(g)}=g_{1}w_{J_{1}(g)}g_{2}w_{J_{2}(g)}\cdots g_{n(g)}
w_{J_{n(g)}(g)}=(g_{1})^{'}(g_{2})^{'}\cdots (g_{n(g)})^{'}$$
By using this formula we can write the first step of our algorithm
(in Proposition 2.2.18) as follows:
$$\Lambda^{\nu}_{[1..n]}(g)=(-1)^{n-n(g)}\Lambda^{\nu}_{\sigma (g)}
|Q^{\nu}(g^{'})|^{2}\Lambda^{\nu}_{J_{1}(g)}((g_{1})^{'})
\Lambda^{\nu}_{J_{2}(g)}((g_{2})^{'})\cdots \Lambda^{\nu}_{J_{n(g)}(g)}
((g_{n(g)})^{'}) \eqno(*)$$
For $g^{''}=g^{'}w_{J(g^{'})}$ we also have the formula $g^{''}=
(g_{1})^{''}(g_{2})^{''}\cdots (g_{n(g)})^{''}$, so the second step of our
algorithm gives
\begin{eqnarray*}
&&\Lambda^{\nu}_{J_{1}(g)}((g_{1})^{'})\Lambda^{\nu}_{J_{2}(g)}
((g_{2})^{'})\cdots \Lambda^{\nu}_{J_{n(g)}(g)}((g_{n(g)})^{'})=\\
&&=(-1)^{n-n(g^{'})}\Lambda^{\nu}_{\sigma ((g_{1})^{'}|J_{1}(g))}
\cdots \Lambda^{\nu}_{\sigma ((g_{n(g)})^{'}|J_{n(g)}(g))}
|Q^{\nu}(g^{''})|^{2}\Lambda^{\nu}_{J(g^{'})}(g^{''})\\
&&=(-1)^{n-n(g^{'})}\Lambda^{\nu}_{\sigma (g^{'}):\sigma (g)}
|Q^{\nu}(g^{''})|^{2}\Lambda^{\nu}_{J(g^{'})}(g^{''})
\end{eqnarray*}
By substituting this into (*) and using the following general
fact (which is immediate from the definition of $Q^{\nu}(g)=\prod
_{a<b,g^{-1}(a)>g^{-1}(b)}Q^{\nu}_{a,b}$):
$$Q^{\nu}(g)Q^{\nu}(g^{'})=Q^{\nu}(g)Q^{\nu}(gw_{J(g)})=Q^{\nu}(w_{J(g)})
=Q^{\nu}(w_{J_{1}(g)})\cdots Q^{\nu}(w_{J_{n(g)}(g)})$$
($\Rightarrow |Q^{\nu}(g)Q^{\nu}(g^{'})|^{2}=Q^{\nu}_{J_{1}(g)}
Q^{\nu}_{J_{2}(g)}\cdots Q^{\nu}_{J_{n(g)}(g)}=Q^{\nu}_{\sigma (g)}$)
we finally obtain the desired formula.

Now we shall state our principal result concerning the inversion of
matrices $A^{(\nu )}$ of the sesquilinear form $(\ ,\ )_{\bf q}$,
defined in 1.3, on the generic weight space ${\bf f}_{\nu}, |\nu |=n$.
\bthm{THEOREM 2.2.20. [INVERSE MATRIX COEFFICIENTS]} Let $\nu $ be a generic weight, $|\nu |=n$. For the
coeficients $\Lambda^{\nu}(g)$ in the expansion
$$[A^{(\nu )}]^{-1}=\sum_{g\in S_{n}}\Lambda^{\nu}(g)\hat{R}_{\nu}(g)$$
we have, with the notations above, the following formulas:\newline
i) If $g\in S_{n}$ is a tree-like permutation of depth $d$, then
$$\Lambda^{\nu}(g)=(-1)^{N(g)}\Lambda^{\nu}_{\sigma (g)}
\Lambda^{\nu}_{\sigma (g^{'}):\sigma (g)}\Lambda^{\nu}_{\sigma (g^{''})
:\sigma (g^{'})}\cdots \Lambda^{\nu}_{\sigma (g^{(d)}):\sigma (g^{(d-1)})}
Q^{\nu}_{\sigma (g^{'})}Q^{\nu}_{\sigma (g^{'''})}\cdots Q^{\nu}_{\sigma
(g^{(d^{'})})}$$
where $\dsty N(g):=\sum_{k=0}^{d}\sum_{I\in \sigma (g^{(k)})}(Card\ I-1), \ d^{'}=
2\left\lfloor\frac{d-1}{2}\right\rfloor+1$\newline
ii) If $g\in S_{n}$ is not tree-like, then $\Lambda^{\nu}(g)=0$.
\ethm
{\bf Proof.} i) follows by iterating our fast algorithm (of Prop.2.2.19).\newline
ii) If $g$ is not tree-like then in the Young sequence of $g$ we encounter
some Young factor which together with its reverse is not splittable,
but then the corresponding $\Lambda^{\nu}_{[..]}(\hbox{the factor})=0$
(c.f. Prop.2.2.11), hence $\Lambda^{\nu}(g)=0$.

Now we give explicit formulas for the inverses of $A^{123}$ and
$A^{1234}$: 
\begin{eqnarray*}
[A^{123}]^{-1}&=&\frac{1}{\bbox_{[1..3]}}\{ \frac{I-Q_{[1..2]}Q_{[2..3]}}
{\bbox_{[1..2]}\bbox_{[2..3]}}(\hat{R}(123)+\hat{R}(321))-\\
&-&\frac{1}{\bbox_{[1..2]}}(\hat{R}(213)+Q_{[1..2]}\hat{R}(312))
-\frac{1}{\bbox_{[2..3]}}(\hat{R}(132)+Q_{[2..3]}\hat{R}(231))\}.\\{}
[A^{1234}]^{-1}&=&\Lambda^{1234}(id)\hat{R}(1234)+\frac{1}{\bbox_{1234}}
\{ -\frac{I-Q_{123}Q_{34}}{\bbox_{12}\bbox_{123}\bbox_{34}}\hat{R}(2134)\\
&-&\frac{I-Q_{123}Q_{234}}{\bbox_{23}\bbox_{123}\bbox_{234}}\hat{R}(1324)
-\frac{I-Q_{12}Q_{234}}{\bbox_{12}\bbox_{34}\bbox_{234}}\hat{R}(1243)
+\frac{1}{\bbox_{12}\bbox_{34}}\hat{R}(2143)+\\
&+&\frac{I-Q_{12}Q_{23}}{\bbox_{12}\bbox_{23}\bbox_{123}}\hat{R}(3214)-
\frac{Q_{12}}{\bbox_{12}\bbox_{123}}\hat{R}(3124)
-\frac{Q_{23}}{\bbox_{23}\bbox_{123}}\hat{R}(2314)\\
&+&\frac{I-Q_{23}Q_{34}}{\bbox_{23}\bbox_{34}\bbox_{234}}\hat{R}(1432)
-\frac{Q_{23}}{\bbox_{23}\bbox_{234}}\hat{R}(1423)
-\frac{Q_{34}}{\bbox_{34}\bbox_{234}}\hat{R}(1342)\} \\
&+&\hbox{(eleven terms obtained by multiplying with $-\hat{R}(4321)$)}.
\end{eqnarray*}
For $\Lambda^{123}(id)$ and $\Lambda^{1234}(id)$ see examples to
Cor.2.2.13.\newline
(Here we abbreviated $Q_{[1..2]}, Q_{[2..4]}$ to $Q_{12}$ (don't
confuse with $Q_{1,2}$), $Q_{234}$ etc.).

Note that $A^{1234}$ is a $24\times 24$ symbolic matrix so the
inversion of such a matrix by standard methods on a computer is almost
impossible (the output may contain hundreds of pages of messy expressions!).

\rem{REMARK 2.2.21.} By using our reduction to the generic case formula 1)
$[A^{(\nu)}]_{\bf ij}^{-1} = \sum_{h\in H} [\tilde
A^{(\tilde\nu)}]_{{\bf\tilde i}, h{\bf\tilde j}}^{-1}$ in 1.7 we can write
also formulas for the inverse matrix coefficients in the case of degenerate
weights $\nu$. E.g. for the inverse of $A^{113}$ (see Example 1.6.4) one gets
$$[A^{113}]^{-1}=\frac1\Delta\left(\begin{array}{ccc}
1 & -(1+q_{11})q_{13}&q_{11}q_{13}^2\\
-q_{31}(1+q_{11}) & (1+q_{11})(1+q_{13}q_{31}) & -(1+q_{11})q_{13}\\
q_{13}^2q_{11} & -q_{31}(1+q_{11}) & 1
\end{array}\right)$$
where
$$\Delta=(1+q_{11})(1-q_{13}q_{31})(1-q_{11}q_{13}q_{31}).$$

\newpage
\section{Applications} 
\subsection{Quantum bilinear form of the discriminant arrangement of
hyperplanes} 

Here we briefly recall the definition of the quantum bilinear form
in case of the configuration ${\cal A}_{n}$ of diagonal hyperplanes
$H_{ij}=H_{ij}^{n}:x_{i}=x_{j}, 1\leq i<j\leq n$ in ${\bf R}^{n}$ (for
general case see [Var]). This arrangement ${\cal A}_{n}$ is also called
the {\em discriminant arrangement} of hyperplanes in ${\bf R}^{n}$.
The {\em domains} of ${\cal A}_{n}$ (i.e connected components of the complement
of the union of hyperplanes in ${\cal A}_{n}$) are clearly of the form
$$P_{\pi }=\{ x\in {\bf R}^{n}| x_{\pi (1)}<x_{\pi (2)}<\cdots <x_{\pi (n)}\}
,\pi \in S_{n}$$
Let $a(H_{ij}^{n})=q_{ij}$ be the {\em weight} of the hyperplane $H_{ij}\in {\cal A}_{n}$, where $q_{ij}$ are given real numbers, $1\leq i<j\leq n$.
Then the {\em quantum bilinear form} $B_{n}$ of ${\cal A}_{n}$ is defined on the
free vector space $M_{n}=M_{{\cal A}_{n}}$ generated by the domains of
${\cal A}_{n}$ by
$$B_{n}(P_{\pi },P_{\tau })=\prod a(H)$$
where the product is taken over all the hyperplanes $H\in {\cal A}_{n}$
which separate $P_{\pi }$ and $P_{\tau }$.
\bthm{PROPOSITION 3.1.1.} We have
$$B_{n}(P_{\pi },P_{\tau })=\prod_{(a,b)\in I(\pi^{-1})\triangle I(\tau^{-1})}
q_{ab}$$
where $I(\sigma )=\{ (a,b)| a<b, \sigma (a)>\sigma (b)\} $ denotes the set of
inversions of $\sigma \in S_{n}$ and $X\triangle Y=(X\setminus Y)\bigcup
(Y\setminus X)$ denotes the symmetric difference of sets $X$ and $Y$.
\ethm
{\bf Proof.}  For each hyperplane $H_{ab}: x_{a}=x_{b}$ we denote by
$H^{+}_{ab}: x_{a}<x_{b}$ and $H^{-}_{ab}: x_{a}>x_{b}$ the corresponding
open half-spaces. Then $H_{ab}$ separates domains $P_{\pi }$ and $P_{\tau }$
if either \newline
1) $P_{\pi }\subset H^{+}_{ab}$ and $P_{\tau }\subset H^{-}_{ab}$ \ \ or \newline
2) $P_{\pi }\subset H^{-}_{ab}$ and $P_{\tau }\subset H^{+}_{ab}$. \newline
In case 1) we have $\pi^{-1}(a)<\pi^{-1}(b)$ and $\tau^{-1}(a)>\tau^{-1}(b)$
i.e $(a,b)\in I(\tau^{-1})\setminus I(\pi^{-1})$. Similary in case 2) we have
$(a,b)\in I(\pi^{-1})\setminus I(\tau^{-1})$. The proof is finished.
\bthm{COROLLARY 3.1.2.} The matrix of the quantum bilinear form $B_{n}$ of
the discriminant arrangements ${\cal A}_{n}=\{ H_{ij}\} $ of hyperplanes in
${\bf R}^{n}$ coincides with the matrix $A^{12\cdots n}=A^{12\cdots n}
({\bf q})$ of the form $(\ ,\ )_{{\bf q}}$ (defined in 1.3), restricted
to the generic weight
space ${\bf f}_{\nu}$, where $I=\{ 1,2,\dots ,n\} ,\nu_{1}=\nu_{2}=\cdots
=\nu_{n}=1$ and where ${\bf q}=\{ q_{ij}\in {\bf R}, 1\leq i, j\leq n,
q_{ij}=q_{ji}\} , q_{ij}=$ the weight of $H_{ij}$ for $1\leq i<j\leq n$.
\ethm

This Corollary enables us to translate all our results concerning matrices
$A^{\nu}, \nu =$ generic, $|\nu |=n$ into results about the quantum bilinear
form $B_{n}$. As an example we shall reinterpret our determinantal formula
given in Theorem 1.9.2. First we recall some definitions and results from
[Var] for the case of the configuration ${\cal C}={\cal A}_{n}$. An {\em
edge} of ${\cal A}_{n}$ is any nonempty intersection of some subset of
hyperplanes of the configuration ${\cal A}_{n}$. The set of all edges of
${\cal A}_{n}$ we denote by ${\cal E}({\cal A}_{n})$. The {\em weight of an edge}
is defined to be the product of the weights of all hyperplanes which contain
the edge. Then the Varchenko's formula (c.f. Theorem (1.1) in [Var])
reads as follows
$$\det B_{n}=\prod_{L\in {\cal E}({\cal A}_{n})}(1-a(L)^{2})^{l(L)}$$
where $a(L)$ is the weight of the edge $L, l(L)$ is the {\em multiplicity
of the edge}, defined in Section 2 of [Var].

In order to state our formula we denote by ${\cal E}^{'}({\cal A}_{n})
\subset {\cal E}({\cal A}_{n})$ the set of those edges which belong to {\em
$k$-equal subspace arrangements} ${\cal A}_{n,k}=\{ x_{i_{1}}=\dots =x_{i_{k}} :
1\leq i_{1}<i_{2}<\cdots <i_{k}\leq n\} ,k\geq 2$, i.e
$${\cal E}^{'}({\cal A}_{n})={\cal A}_{n,n}\cup {\cal A}_{n,n-1}\cup
\cdots \cup{\cal A}_{n,2}$$
\bthm{THEOREM 3.1.3.} The determinant of the quantum bilinear form $B_{n}$
of the discriminant arrangement ${\cal A}_{n}$ is given by the formula
$$\det B_{n}=\prod_{L\in {\cal E}^{'}({\cal A}_{n})}(1-a(L)^{2})^{l(L)}$$
where for $L=\{ x_{i_{1}}=x_{i_{2}}=\cdots =x_{i_{k}}\} \in {\cal A}_{n,k}
\subset {\cal E}^{'}({\cal A}_{n})$ we have
$$a(L)=\prod_{1\leq a<b\leq k}q_{i_{a}i_{b}}, l(L)=(k-2)!(n-k+1)!$$
\ethm
{\bf Proof.} In Theorem 1.9.2 we set $I=\{ 1,2,\dots ,n\} ,\nu_{1}=\nu_{2}
=\cdots \nu_{n}=1$ and for $\mu\subseteq \nu $ interpreted as the set
$\{ i_{1},i_{2},\dots ,i_{k}\} $ we obtain $\Box_{\mu}=1-{q}_{\mu}
=1-\prod_{i\neq j\in \mu}q_{ij}=1-(\prod_{1\leq a<b\leq k}q_{i_{a}i_{b}})
^{2}$ (here $q_{ij}$ are real!). But $a(L)=\prod_{H\supseteq L}a(H)
=\prod_{1\leq a<b\leq k}a(H_{i_{a}i_{b}})=\prod_{1\leq a<b\leq k}q_{i_{a}i_{b}}
$, so $\Box_{\mu}=1-a(L)^{2}$. Note that $|\mu|=k, |\nu |=n$.
Now the proof follows by Corollary 3.1.2.

Note that our formula for $\det B_{n}$ is more explicit then Varchenko's
formula, and in particular we conclude that the multiplicity
$l(L)=0$ for all $L\in {\cal E}({\cal A}_{n})\setminus{\cal E}^{'}({\cal
A}_{n})$.

\subsection{Quantum groups} 

We shall adopt the notations used in [SVa]. Fix the following data:\newline
a) a finite dimensional complex vector space ${\goth h} $\newline
b) a non-degenerate symmetric bilinear form $(\ ,\ )$ on ${\goth h} $ \newline
c) linearly independent covectors ("simple roots") $\alpha_{1},\dots
,\alpha_{n}\in {\goth h}^{*}$\newline
d) a non-zero complex number $\kappa $. \newline
Let $b: {\goth h} \rightarrow {\goth h}^{*}$ be the isomorphism induced by
$(\ ,\ )$. We transfer the form $(\ ,\ )$ to ${\goth h}^{*}$ using $b$.
Put $b_{ij}=(\alpha_{i},\alpha_{j})$; $B=(b_{ij})\in Mat_{r}({\bf C}),
h_{i}=b^{-1}(\alpha_{i})\in {\goth h} $.

Put $q=\exp (2\pi i/\kappa )$; for $a\in {\bf C}$ put $q^{a}=\exp (2\pi ia/
\kappa )$.

Let $U_{q}{\goth g} =U_{q}{\goth g} (B)$ be the ${\bf C}$-algebra generated
by elements $e_{i}, f_{i}, i=1,\dots ,n$ and the space ${\goth h} $, subject
to relations
\begin{eqnarray*}
[h,e_{i}]&=&\alpha_{i}(h)e_{i}; [h,f_{i}]=-\alpha_{i}(h)f_{i}\\{}
[e_{i},f_{j}]&=&(q^{h_{i}/2}-q^{-h_{i}/2})\delta_{ij}\\{}
[h,h^{'}]&=&0
\end{eqnarray*}
for all $i,j=1,\dots ,n; h,h^{'}\in {\goth h} $.

The comultiplication $\triangle : U_{q}{\goth g} \rightarrow U_{q}{\goth g}
\hat{\otimes }U_{q}{\goth g} $ is given by $\triangle (h)=h\otimes 1 +
1\otimes h, \triangle (f_{i})=f_{i}\otimes q^{h_{i}/4} + q^{-h_{i}/4}
\otimes f_{i}, \triangle (e_{i})=e_{i}\otimes q^{h_{i}/4} + q^{-h_{i}/4}
\otimes e_{i}$.

The counit $\epsilon :U_{q}{\bf g} \rightarrow {\bf C}$ is defined by
$\epsilon (f_{i})=\epsilon (e_{i})=\epsilon (h)=0$ and the antipode
$A: U_{q}{\goth g} \rightarrow U_{q}{\goth g} $ by $A(h)=-h,
A(e_{i})=-q^{b_{ii}/4}e_{i}, A(f_{i})=-q^{-b_{ii}/4}f_{i}$. We denote by
$U_{q} {\goth n}_{-}$ (resp. $U_{q}{\goth n}_{+}, U_{q}{\bf h} $) subalgebras
generated by $f_{i}$ (resp. $e_{i}, h\in {\goth h} $), $i=1,\dots ,n$.
$U_{q}{\goth n}_{\pm}$ are free. We have $U_{q}{\goth g} =U_{q}{\goth n}_{-}\cdot
U_{q}{\goth h} \cdot U_{q} {\goth n}_{+}$.

For $\lambda =(k_{1},\dots ,k_{n})\in {\bf N}^{n}$ put
$$(U_{q}{\goth n}_{-})_{\lambda }=\{ x\in U_{q}{\goth n}_{-}| [h,x]=-\sum
k_{i}\alpha_{i}(h)x \forall h\in {\goth h} \} $$
We have $U_{q}{\goth n}_{-}=\oplus_{\lambda}(U_{q}{\goth n}_{-})_{\lambda}$.
\rem{\bf Contravariant forms.}  There exist a unique symmetric bilinear
form $S$ on $U_{q}{\goth n}_{-}$ satisfying
$$S(1,1)=1, S(f_{i}x,y)=S(x,g_{i}y)$$
where $g_{i}: (U_{q}{\goth n}_{-})_{\lambda}\rightarrow (U_{q}{\goth n}_{-})
_{\lambda^{'}_{i}}, i=1,\dots ,n; \lambda^{'}_{i}=(k_{1},\dots ,k_{i}-1,\dots
,k_{n})$ are the operators acting on $f_{J}=f_{j_{1}}\cdots f_{j_{n}}\in
(U_{q}{\goth n}_{-})_{\lambda}$ as follows:
$$g_{i}(f_{J}):=\sum_{p:j_{p}=i}q^{\sum_{l<p}b_{ij_{l}}/4-\sum_{l>p}
b_{ij_{l}}/4}f_{j_{1}}\cdots \hat{f}_{j_{p}}\cdots f_{j_{n}}.$$
If the weight $\lambda =(k_{1},\dots ,k_{n})=(1,1,\dots ,1)$, then for
$f_{I}=f_{i_{1}}f_{i_{2}}\cdots f_{i_{n}}, f_{J}=f_{j_{1}}f_{j_{2}}
\cdots f_{j_{n}}\in (U_{q}{\goth n}_{-})_{\lambda}$ is given explicitly by
$$S(f_{I},f_{J})=q^{(\sum_{k<l}\pm b_{i_{k}i_{l}})/4}$$
where in the sum we take $+b_{i_{k}i_{l}}$ if
$\sigma (k)>\sigma (l)$ and $-b_{i_{k}i_{l}}$ otherwise.

Here $\sigma =\sigma (I,J)\in S_{n}$ is the unique permutation such
that $j_{p}=i_{\sigma (p)}$ for all $p$.
\bthm{THEOREM 3.2.1.} The determinant of the contravariant form $S$ on the
weight space $(U_{q}{\goth n}_{-})_{(1,1,\dots ,1)}$ is given by the following
formula
\begin{eqnarray*}
&&\det S|_{(U_{q}{\bf n}_{-})_{(1,1\dots ,1)}}\\
&&=q^{-\frac{n!}{4}\sum_{1\leq k<l\leq
n}b_{kl}}\prod_{m=2}^{n}\prod_{1\leq i_{1}<\cdots <i_{m}\leq n}(1-q^{\sum_{
1\leq k<l\leq m}b_{i_{k}i_{l}}})^{(m-2)!(n-m+1)!}\\
&&=\prod_{m=2}^{n}\prod_{1\leq i_{1}<\cdots <i_{m}\leq n}(q^{-\frac{1}{2}
\sum_{1\leq k<l\leq m}b_{i_{k}i_{l}}}-q^{\frac{1}{2}\sum_{1\leq k<l\leq m}
b_{i_{k}i_{l}}})^{(m-2)!(n-m+1)!}
\end{eqnarray*}
\ethm
{\bf Proof.} By factoring out from the matrix $S(f_{I},f_{J})$
the factor $q^{-\frac{1}{4}\sum_{1\leq k<l\leq n}b_{kl}}$ we get a matrix which
(up to permutation of rows and columns) coincides with the matrix $A^{12\cdots
n}({\bf q})$, where ${\bf q}=\{ q_{ij}\} $, $q_{ij}:=q^{-\frac{1}{2}b_{ij}}$.
Then we apply Theorem 1.9.2 and the result follows.

\newpage\thispagestyle{empty}

\end{document}